\documentclass[apj]{emulateapj}

\usepackage{epsfig}
\usepackage{graphicx}
\usepackage{amsmath}
\usepackage{natbib}

\shorttitle{The Galactic Winds of HCG16  }
\shortauthors{Vogt et al.}

\begin{document}

\title{Galaxy Interactions in Compact Groups I : The Galactic Winds of HCG16}

\author{Fr\'ed\'eric P.A. Vogt\altaffilmark{1}, Michael A. Dopita \altaffilmark{1}, Lisa J. Kewley \altaffilmark{1}}
\email{fvogt@mso.anu.edu.au}

\altaffiltext{1}{Mount Stromlo Observatory, Research School of Astronomy and Astrophysics, The Australian National University, Cotter Road, Weston Creek, ACT 2611, Australia.}

\begin{abstract}
Using the WiFeS integral field spectrograph, we have undertaken a series of observations of star-forming galaxies in Compact Groups. In this first paper dedicated to the project, we present the analysis of the spiral galaxy NGC838, a member of the Hickson Compact Group 16, and of its galactic wind. Our observations reveal that the wind forms an asymmetric, bipolar, rotating structure, powered by a nuclear starburst. Emission line ratio diagnostics indicate that photoionization is the dominant excitation mechanism at the base of the wind. Mixing from slow shocks (up to 20\%) increases further out along the outflow axis. The asymmetry of the wind is most likely caused by one of the two lobes of the wind bubble bursting out of its H{\footnotesize I} envelope, as indicated by line ratios and radial velocity maps. The characteristics of this galactic wind suggest that it is caught early (a few Myr) in the wind evolution sequence. The wind is also quite different to the galactic wind in the partner galaxy NGC839 which contains a symmetric, shock-excited wind. Assuming that both galaxies have similar interaction histories, the two different winds must be a consequence of the intrinsic properties of NGC838 and NGC839 and their starbursts.\\
\end{abstract}


\section{Introduction}\label{Sec:intro}

Galaxy evolution is influenced by a wide range of phenomena associated with galaxy interactions. These phenomena include the onset or quenching of star formation, the triggering of large scale galactic winds, gas stripping, tidal perturbations, or the activation and feeding of AGNs (Active Galactic Nuclei) and their associated feedback. Compact groups of galaxies (CGs) are dense environments prone to intense and repeated galaxy interactions \citep[][]{Coziol07}. In these groups, gas-rich, star forming galaxies evolve into gas-poor, quiescent objects. Galaxies may be first processed in CG-style environments, before being transformed within filaments and larger clusters \citep[][]{Moran07,Cappellari11,Mahajan12}. The mid-infrared (mid-IR) colors of galaxies in CGs also suggest that galaxy evolution is accelerated in CG environments compared with the field \citep[][]{Walker10}. 

The first catalog of 100 low redshift CGs was compiled by \cite{Hickson82b,Hickson82a} using the Palomar Observatory Sky Survey red prints. The gravitationally bounded nature of the majority of these Hickson Compact Groups (HCGs) was later confirmed using (1) spectroscopic observations \citep[][]{Hickson92} and (2) X-ray observations detecting emission from hot gas surrounding the group members \citep[][]{Ponman96}. Recently, \cite{Iovino02} extended the survey and compiled a list of 121 Southern Compact Groups (SCGs). CGs usually contain 5 to 15 members, and have a radius of a few hundred kpc.

CGs, and HCGs especially, have been subject to close scrutiny ever since their discovery. Radio observations of HCGs by  \cite{Verdes-Montenegro01} revealed an overall deficiency of H{\footnotesize I} gas of the order of 60\%. \cite{Martinez-Badenes12} confirmed this result and using CO lines detected an excess of molecular hydrogen in spiral HCG galaxies of the order of 50\%. \cite{Martinez10} found that HCGs contain a much higher fraction of low-luminosity AGN than the field or cluster environments. Yet, the overall star formation rate (SFR) of galaxies in HCGs is not significantly different from galaxies in the field \citep[][]{Verdes-Montenegro98,Martinez10,Bitsakis10, Bitsakis11}. \cite{Rosa07} also observed that the HCG stellar population is on average older and more metal-poor than the field. These observations suggest that interactions-prone CGs quench star formation in the long term. The causal link between galaxy interactions and enhanced star formation activity is well established \citep[][]{Barton00,Bournaud11, Alonso12}, but (1) these starburst phases are short lived, and (2) not all interactions trigger a starburst episode \citep[][]{Matteo07,Matteo08}. The competing role between one-on-one interactions and the local environment of galaxies towards the type (and intensity) of nuclear activity has been directly observed by \cite{Sabater13} in a statistical analysis of galaxies in the Sloan Digital Sky Survey \citep[SDSS;][]{York00}.

In a so-called \emph{wet} merger\footnote{for gas-rich galaxies, hydrodynamics is important, hence the term \emph{wet} from hydro, the Greek root for water.} \citep[e.g.][]{Joseph85}, the galaxy's first encounter gives rise to tidal disturbances, often seen in the form of tidal tails and non-regular rotation curves. These tidal disturbances subsequently trigger a gas-infall towards the nuclei of the interacting galaxies. This strong and rapid infall induces shock waves that ionize the in-falling gas. The ionized gas cools, first via X-ray emission, later by optical and UV line emission, and finally by the excitation of molecular hydrogen formed in the warm, recombined post-shock gas. Molecular hydrogen excitation is a more efficient cooling pathway than X-ray emission in certain systems. Stephan's Quintet is a prime example \citep{Cluver10}. The extended formation of molecular hydrogen (H$_2$) creates large and dense clouds of warm H$_2$. Shocks further induce gravitational instabilities in these gas clouds, which leads to strong episodes of massive star formation.

Theoretical models that incorporate the physics of the Interstellar Medium (ISM) over small scales ($<$1 pc) are lacking. Yet, they are critical to gain a detailed understanding of the physics of galaxy evolution in environments such as CGs. For example, \cite{Teyssier10} present simulations with a resolution of 12 pc. They conclude that gas fragmentation into large clouds is the dominant star formation process. This mechanism can only be seen if the densest phases of the ISM are resolved.

One of the main limitations of our current understanding of galaxy evolution is the type of instrumentation that has been available until recently. Galaxies are complex systems that need to be fully resolved both spatially and spectrally, and therefore observations with single fibers or with long-slit spectrographs are inadequate. Fortunately, Integral Field Spectrographs (IFS) now provide astronomers with the optimum tool to tackle galaxy evolution at the sub-kpc level in the local Universe. The unintended discovery of a galactic wind during the first commissioning run of the Sydney-AAO Multi-object Integral Field spectrograph \citep[SAMI;][]{Croom12} is one example of the unique potential of IFS for galaxy surveys \citep{Fogarty12} .

We are undertaking a series of IFS observations of star forming galaxies in CGs using the WiFeS instrument \citep[][]{Dopita07} on the Australian National University (ANU) 2.3m telescope at Siding Spring Observatory. Because of its large field of view (25$\times$38 square arcseconds), WiFeS is among the very few IFS currently capable of observing the entire extent of CG galaxies in a reasonable amount of time. Our targets, which usually span over 1 arcminute on sky, can be mapped completely with just a few pointings. With this series of observations, we aim to investigate galaxy interactions, the formation of galactic winds and the onset of star formation episodes at kpc scales. This project is complementary to recent WiFeS observing campaigns of isolated dwarf galaxies \citep[][Nicholls et al. 2013, in preparation]{Nicholls11}, strong mergers \citep{Rich11}, LIRGs, \citep[Luminous Infrared Galaxies;][]{Rich12} and galaxies in cluster environments \citep[][]{Farage10,Farage12,Merluzzi12}.  

The \emph{low-redshift \& high resolution} approach of our project is complementary to other ongoing efforts to improve our understanding of galaxy evolution at high redshift, for example by using gravitational lensing to push the redshift limit of spatially resolved targets \citep[e.g.][]{Jones10,Yuan11}. Our study provides an important benchmark sample for these high redshift studies. We select galaxies based on their mid-IR colors using WISE images \citep[][]{Wright10}, UV color using Galex images \citep[][]{Martin05}, and optical color and morphology using Digital Sky Survey (DSS) images. This selection focuses the sample on galaxies with ongoing star formation. We use tidal features in the optical images to detect ongoing interactions.

This series of observations is not designed nor intended as a survey. While we plan to target galaxies with a wide range of nearest-neighbour distances, located in a wide range of CGs sizes, our focus is on the detailed analysis of these galaxies and the physics at work over their whole spatial extent. 

In this article, we present the results of our IFS observations of the galaxy NGC838 and reveal the true nature of its galactic wind. The paper is organized as follows. We introduce HCG16 and NGC838 in Section~\ref{sec:838}. Our observations and data reduction procedure are described in Section~\ref{sec:obs}. Our method for extracting emission line fluxes and velocities is detailed in Section~\ref{sec:analysis}.  Our results are shown in Section~\ref{sec:results}. We discuss their implications in Section~\ref{sec:discussion} and summarise our conclusions in Section~\ref{sec:conclusion}. In this paper, we assume $H_0=71$ km s$^{-1}$ Mpc$^{-1}$, $\Omega_M=0.27$ and $\Omega_V$=0.73, following the 7 years WMAP results \citep[][]{Larson11}.

\section{NGC838 and HCG16}\label{sec:838}

In the initial definition of \cite{Hickson82a}, HCG16 was catalogued as containing 4 galaxies, with an on-sky diameter of 1.6 arcminutes. Based on spectroscopic follow-up observations \citep[][]{Carvalho97}, the total number of member galaxies has since increased to 7. Evidences suggest that HCG16 may be part of a larger structure \citep[][]{Ribeiro98}, where a core of 5 galaxies is surrounded by a halo of several others yet to fall deeper in the potential well. HCG16 is one of the most active of the HCGs \citep{Ribeiro96}. In Table~\ref{table:hcg16}, we list the different group members and their activity type. Based on the H{\footnotesize I} structure of this group, \cite{Verdes-Montenegro01} selected HCG16 as a typical example of the second phase of their 5-step evolution sequence for the H{\footnotesize I} distribution in Compact Groups, where 30\%-60\% of the total H{\footnotesize I} gas originally located in the halo of galaxies is dispersed in tidal features. In their radio map of the group, a clear H{\footnotesize I} bridge is detected between the central 4 galaxies (NGC835,833,838,839) and NGC848. This bridge suggests that NGC848 may be an intruder galaxy that collided with HCG16 in a scenario similar to the archetypical Stefan's Quintet \citep[a.k.a. HCG92, see][]{Sulentic01,Appleton06}.

\begin{table*}[htb!]
\begin{center}\caption{Galaxy members of HCG16 and their activity}\label{table:hcg16}
\begin{tabular}{c c c c c p{3.5cm}}
\tableline\hline
Name & Alt. name\tablenotemark{a} & R.A.\tablenotemark{b} & Dec.\tablenotemark{b} & cz\tablenotemark{b} & Activity\tablenotemark{c}\\
 & & (J2000) &(J2000) & (km s$^{-1}$) & \\
\hline
NGC835 & HCG16-a/1 & 2$^\text{h}$09$^\text{m}$24.6$^\text{s}$ & -10$^\circ$08'09" & 4073 & Seyfert 2, starburst\\
NGC833 & HCG16-b/2 & 2$^\text{h}$09$^\text{m}$20.8$^\text{s}$ & -10$^\circ$07'59" & 3864 & LINER\tablenotemark{d}, starburst\\
NGC838 & HCG16-c/4 & 2$^\text{h}$09$^\text{m}$38.5$^\text{s}$ & -10$^\circ$08'48" & 3851 & starburst\\
NGC839 & HCG16-d/5 & 2$^\text{h}$09$^\text{m}$42.9$^\text{s}$ & -10$^\circ$11'03" & 3874 & LINER\tablenotemark{d}, starburst\\
NGC848 & HCG16-3 & 2$^\text{h}$10$^\text{m}$17.6$^\text{s}$ & -10$^\circ$19'17" & 3989 & starburst \\
KUG 0206-105 & HCG16-6 & 2$^\text{h}$09$^\text{m}$06.0$^\text{s}$ & -10$^\circ$19'13" & 3972 & starburst \\
- & HCG16-10 & 2$^\text{h}$08$^\text{m}$36.8$^\text{s}$ & -09$^\circ$56'16" & 4027 & no emission lines detected\\
\tableline
\end{tabular}
\tablenotetext{1}{ In \cite{Hickson82a,Carvalho97}. The redshifts of HCG16-7,8,9,12,... rule out these galaxies as being members of HCG16.}
\tablenotetext{2}{ R.A., Dec. and heliocentric velocity extracted from NED (Nasa Extragalactic Database).}
\tablenotetext{3}{ See \cite{Ribeiro96, Turner01}.}
\tablenotetext{4}{ Low-Ionization Nuclear Emission Region}
\end{center}
\end{table*}

\begin{figure}[htb!]
\centerline{\includegraphics[scale=0.75]{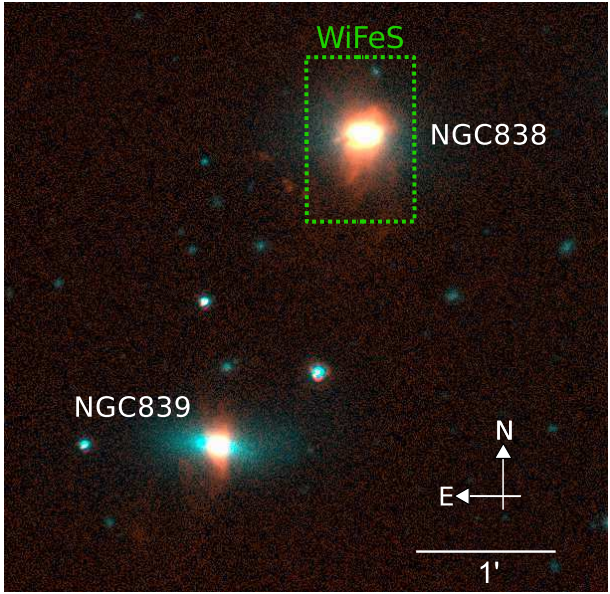}}
\caption{H$\alpha$ (red in the online version) and R-band (blue in the online version) image of NGC838 and NGC839, from the SINGG survey \citep[][]{Meurer06}. The dotted square delineate our WiFeS observation footprint, covering the large majority of the H$\alpha$ emission around NGC838.}\label{fig:singg}
\end{figure}

\begin{figure}[htb!]
\centerline{ \includegraphics[scale=0.28]{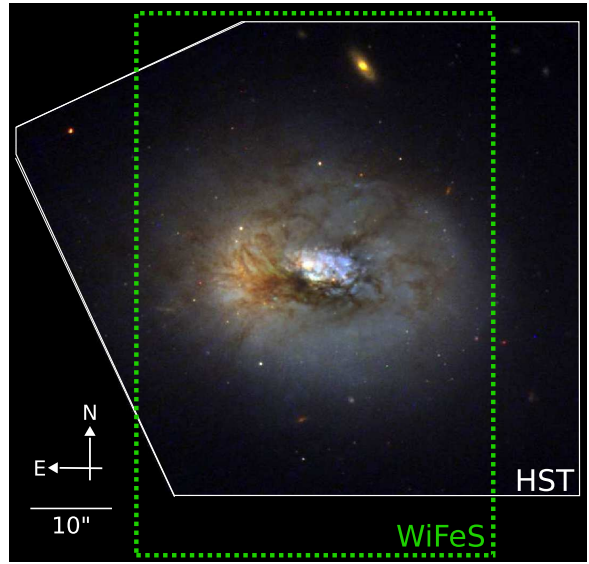}}
\caption{ Wide-Field Planetary Camera 2 (WFPC2) image of NGC838 taken with the Hubble Space Telescope (HST, P.Id.: 10787, P.I.: J.Charlton). Numerous dense dust lanes can be easily identified. The dotted square represents the footprint of our WiFeS observations. This picture is a combination of 2 different images with different contrast settings showing both the bright core and the faint outskirts of the galaxy. The original images were obtained from the Hubble Legacy Archive. A color version of this figure is available in the online version of the article.}\label{fig:hst}
\end{figure}

NGC838 was part of the SINGG survey \citep[Survey for Ionization in Neutral-Gas Galaxies,][]{Meurer06} consisting of H$\alpha$ and R-band imaging for 468 galaxies selected in the HIPASS survey \citep[H{\footnotesize I} Parkes All Sky Survey,][]{Barnes01}. The SINGG image of NGC838 and NGC839 is shown in Figure~\ref{fig:singg}, where the green dotted-line rectangle denotes the footprint of our WiFeS observations. Both NGC838 and NGC839 have an associated extended H$\alpha$ emission region: a narrow and regular structure for NGC839, and an asymmetric, extended structure for NGC838. The galactic wind nature of the H$\alpha$ emission in NGC839, powered by an ongoing starburst, was confirmed by \cite{Rich10}. In Figure~\ref{fig:hst}, we show the HST WFPC2 image of NGC838, and overlay the footprint of our WiFeS observations as a dotted green rectangle. This image clearly shows a nuclear starburst region surrounded by a dense system of filaments of dust and, presumably, molecular gas. At NGC838's distance of $\sim$55 Mpc, 1" corresponds to 0.27 kpc.

\section{Observations and data reduction}\label{sec:obs}

NGC838 was observed with the WiFeS instrument \citep{Dopita07,Dopita10} on the ANU 2.3m telescope at Siding Spring Observatory in October 2011. WiFeS is a dual-beam, image slicer integral field spectrograph with a spatial resolution of 1$\times$0.5 square arcseconds and a spectral resolution of 3000 or 7000 depending on the chosen grating, with a field of view of 25$\times$38 square arcseconds. We have used the B3000 and R7000 gratings with the RT560 dichroic in our observations resulting in a complete wavelength coverage from $\sim$3800{\AA} to $\sim$7000\AA. Our blue spectra have a resolution of 3000 to cover all interesting emission lines blueward of 5700{\AA} and allow for line diagnostic analysis, while the red spectra have a resolution of 7000 to allow for a detailed study of the velocity structure of the H$\alpha$ emission line. We used the 1$\times$2 binning mode, so that each spaxel in the final data cube covers a 1$\times$1 square arcseconds area on-sky.

\begin{figure}[htb!]
\centerline{\includegraphics[scale=0.5]{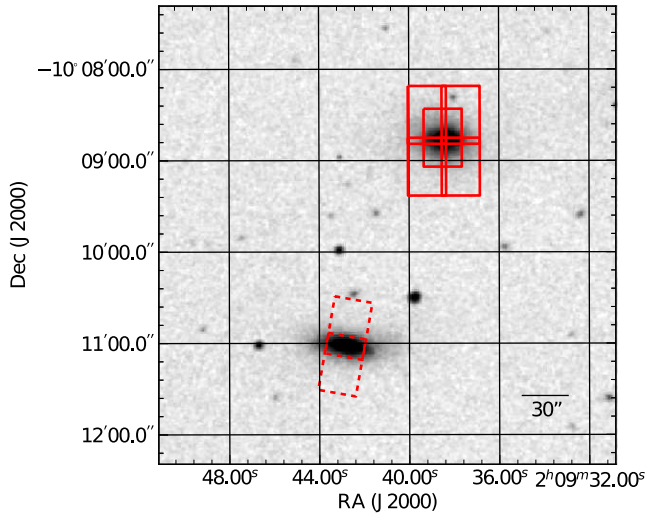}} 
\caption{Red band DSS-2 image of NGC838 (top right) and NGC839 (bottom left). The rectangles (in red in the online version) delineate our WiFeS observation footprint. The dashed-line rectangles show the WiFeS observation footprints of \cite{Rich10} for comparison. The DSS-2 image is copyright (c) 1993-2000 by the AAO and AURA.}\label{fig:finding_chart}
\end{figure}

Four WiFeS pointings were necessary to map NGC838 and its extended H$\alpha$ emission. A fifth field, centred on the galaxy, was observed to assist in the photometric calibration of the different fields used in the final mosaic. In Figure~\ref{fig:finding_chart}, we show the footprint of our WiFeS observations superposed on a red band image of NGC838 from the \emph{Second Digitized Sky Survey} (DSS-2)\footnote{ obtained from the European Southern Observatory Online Digitized Sky Survey Server}, along with the NGC839 observations footprint of \cite{Rich10}. Each field was observed 4 times, 1400 seconds each, resulting in 5600 seconds on target for each pointing. A 700 second sky exposure was interleaved between each science exposure, while telluric and flux standard stars were observed throughout the night. The mosaic took 3.5 nights to complete, none of which was perfectly photometric. Seeing conditions ranged from 1.5" for the bottom left field, 1.8-2" for the central and bottom right field, and 2-2.5" for the top two fields. Data cubes acquired under better seeing conditions were not degraded to match the worst observed seeing of 2.5". Each 4 exposures of a given field (shown in Figure~\ref{fig:finding_chart}) were acquired close enough in time to ensure similar seeing conditions. Because we do a spaxel per spaxel analysis, subsequently degrading one field's spatial resolution to match the spatial resolution of the others is not required and would only result in data of poorer quality.

The data was reduced using the dedicated WiFeS IRAF pipeline (Version of 2011, December; see \cite{Dopita10} for a complete description). Among some other minor updates, this version accounts for the fact that both the blue and the red WiFeS cameras are now read out through one amplifier only (instead of four, as previously).

Each science exposure was reduced individually and converted into a bias subtracted, flat-fielded, flux and wavelength calibrated data cube. Bias subtraction was performed using a technique which accounts for the known temporal variations of the WiFeS biases \citep[][]{Rich10}. Bias frames are acquired throughout the night. For a given science frame, a smooth plane is fitted to the closest acquired bias frame. Using these reconstructed bias frames instead of a standard master bias minimises the errors in bias subtraction. Telluric corrections were applied to the red spectra only. The blue and red spectra are joined at 5700{\AA}, with the blue spectra multiplied by a correction factor of the order of 1.15 to account for a difference in the flux levels at this wavelength. 

We have used the IRAF task \emph{imcombine} to assemble the final mosaic. The reduction pipeline produced 10 reduced cubes (5 red and 5 blue) on the same wavelength grid. No further interpolation along the spectral axis was necessary when the 10 cubes were combined. We limited ourselves to integer spatial shifts, both for simplicity and due to the average seeing conditions during most of our observations. Because header coordinates cannot be trusted to the required accuracy of $\sim$1", we implemented a custom Python program which uses the spatial flux correlations to best match the integrated H$\alpha$ emission in the various overlap regions between fields. The final mosaic is 47 $\times$ 65 square arcseconds in size, and contains 3055 spectra. 

\section{Data analysis}\label{sec:analysis}

\subsection{Stellar templates and emission line fitting}
The underlying gas physics in NGC838 can be understood using the emission line structures and intensities in the data cube. Extracting this information from each of the 3055 spectra contained within our final mosaic is conducted in a two step process. First, the underlying stellar continuum emission needs to be accurately removed. This step is especially critical if emission lines coincide with absorption features, which is the case for (but not only) the H$\alpha$ and H$\beta$ emission lines in NGC838. The second step requires a careful fitting of the sometimes complex line structures to accurately account for the total line flux as well as to obtain an accurate description of the underlying gas kinematics. 

Performing this complex analysis on thousands of spectra is time consuming and requires an automated procedure. Here, we use the dedicated IDL program UHSPECFIT based on stellar fitting routines by \cite{Moustakas06}. The code was originally developed by \cite{Zahid11} and \cite{Rupke10}, and later adapted by \cite{Rich10} to be compatible with WiFeS data. UHSPECFIT is designed to handle the two different spectral resolutions in our data. The stellar continuum is fitted using a combination of stellar templates from \cite{Gonzales-Delgado05}. We are only interested in an accurate removal of the underlying stellar contribution, and therefore limited ourselves to the solar metallicity, Geneva isochrones models, similarly to \cite{Rupke10}. A careful analysis of the underlying stellar population based on a careful stellar population model fitting would require higher S/N spectra, and is outside of the scope of this paper. 

The principal emission lines fitted are H$\alpha$, H$\beta$, [O {\footnotesize III}] $\lambda\lambda$4959,5007, [N {\footnotesize II}] $\lambda$6583, [S {\footnotesize II}] $\lambda\lambda$6716,6731 and [O {\footnotesize I}] $\lambda$6300. Multiple gaussian components are fitted to the continuum-subtracted observed emission lines. The velocity and velocity dispersion of the different components are assumed to be identical for every emission line in a given spectrum. Our line fitting procedure can be summarized mathematically as follows : $n$ emission lines for any given spaxel are described by $n\times m + m + m$ parameters in total ; $n\times m$ intensities, $m$ velocities and $m$ velocity dispersions, where $m\in$[1,2,3] is the number of component associated with a given spaxel. For each spaxel, gaussian components are classified based on their velocity dispersion, from the lowest (component 1, or $c1$) to the highest (component 3, or $c3$).

If most spectra with S/N$>$5 display asymmetric line profiles, only a few have more than 2 components clearly identifiable. Each spectrum with S/N(H$\alpha$)$>$3 is fitted successively with a combination of 1, 2 and 3 gaussian components. We visually inspected every spectrum to determine the number of components that provides the best fit to each spectrum. Only 14 spaxels required 3 gaussian components to provide an adequate fit to the emission line profiles. All other spectra could be well fitted by one or a combination of two gaussian components.

\begin{figure}[htb!]
\centerline{\includegraphics[scale=0.6]{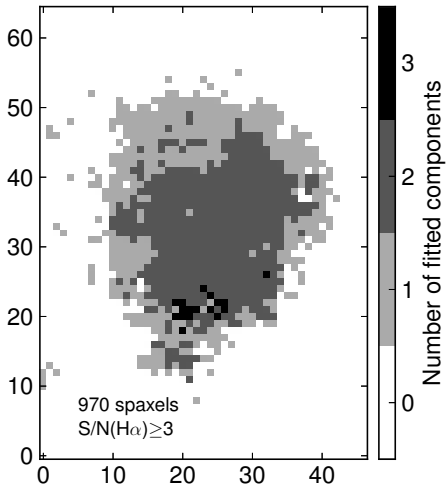}}
\caption{Number of component required to best fit each spectra in the data cube with S/N(H$\alpha$)$\geq$3. Every spectrum was visually inspected to select the number of components best fitting its emission line structures.}\label{fig:components}
\end{figure}

We implemented a simple f-test routine in IDL to ensure that the decrease in the fit residual associated with an additional gaussian component is enough to justify adding a component to the fit \citep[e.g.][]{Westmoquette07}. We adopted a false-rejection probability of 0.05. In most cases, the f-test outcome was consistent with our visual selection, except for the spaxels located towards the center of NGC838, where the automated f-test indicated that three components may be justified. However, (1) two of these components have a width similar to our instrumental spectral resolution and are essentially unresolved, and (2) fitting with only two components provides a fit residual within the noise level of the data in every case, so we favoured our visual selection over the f-test outcome in this region. In Figure~\ref{fig:components}, we show the number of fitted gaussian components used to describe the emission lines structure for the 970 spaxels in our data where S/N(H$\alpha$)$\geq$3. The x and y-axis denote spaxels. A single spaxel corresponds to 1" or 0.27 kpc at the distance of NCG838.

\begin{figure*}[htb!]
\centerline{\includegraphics[scale=0.4]{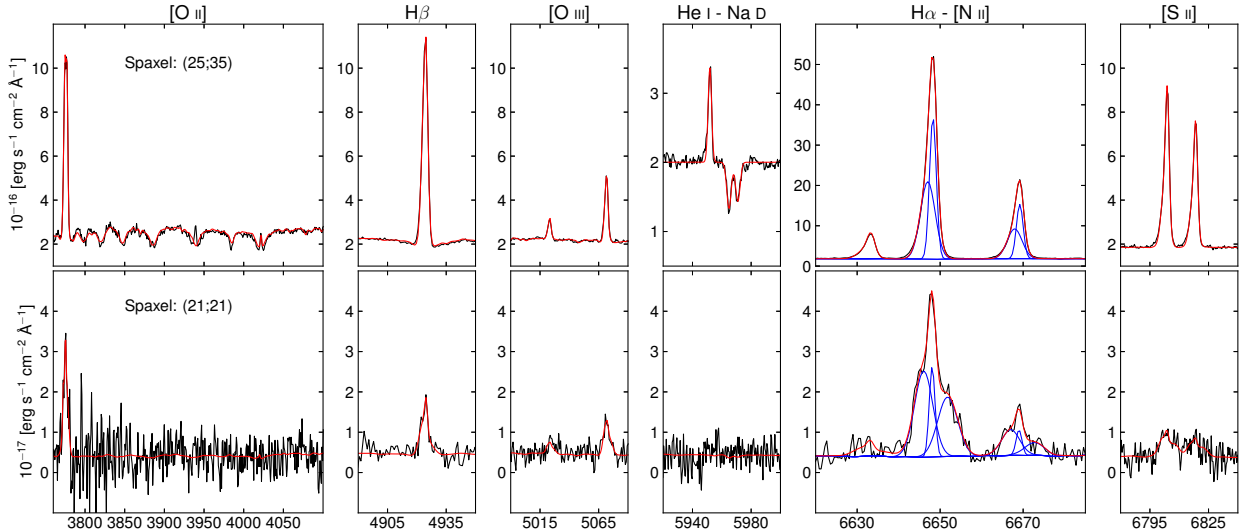}}
\caption{Two examples of spectra (light-grey lines, in black in the online version) and their associated best-fit (black lines, in red in the online version). In the H$\alpha$ - [N {\footnotesize II}] panel, the individual gaussian components used to fit the emission line are shown in dark-grey (blue in the online version). Respectively 2 and 3 components are required to fit the upper and lower spectra. Every column is labelled with the dominant emission line(s) it contains. The quoted wavelengths are the observed ones.}\label{fig:fit}
\end{figure*}

In Figure~\ref{fig:fit}, we show two fit examples that require 2 (upper panel) and 3 (lower panel) gaussian components. In the H$\alpha$ - [N {\footnotesize II}] panel, we also show the different gaussian components used to fit the H$\alpha$ and N[{\footnotesize II}] $\lambda$6583 line profiles. 

The same number of components, different for every spaxel, has been used in the fits to all emission lines, blue or red. Because our spectra have both a high and a low resolution region (7000 in the red, compared to 3000 in the blue), multiple components visible in emission lines located in the high resolution regions are not always as clearly visible in the emission lines blueward of the stitch wavelength at 5700{\AA}. We have visually ensured that all red and blue emission line fit residuals are at the noise level of the data. However, individual gaussian components for the emission lines blueward of 5700{\AA} cannot be trusted because of their reduced S/N, and we will not use these individually.

\subsection{Galactic and extragalactic reddening}\label{sec:red}

Every spectrum in the final mosaic was corrected for galactic extinction using the \cite{Schlafly11} recalibration of the \cite{Schlegel98} extinction map based on dust emission measured by COBE/DIRBE and IRAS/ISSA. The recalibration assumes a \cite{Fitzpatrick99} reddening law with $R_V$=3.1 and a different source spectrum than \cite{Schlegel98}. 

\begin{figure}[htb!]
\centerline{\includegraphics[scale=0.6]{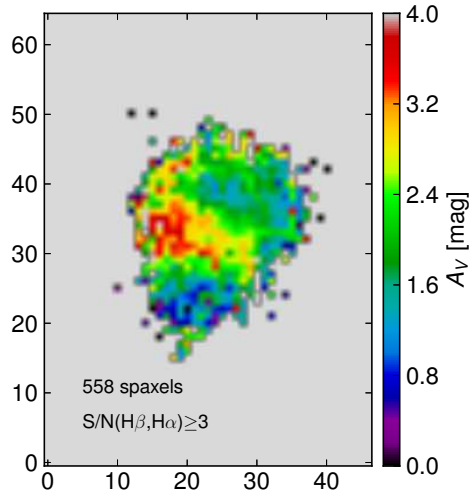}}
\caption{Reddening map of NGC838 in A$_V$ magnitudes based on the H$\alpha$ to H$\beta$ line ratios for spaxels with S/N(H$\alpha$,H$\beta$)$\geq$3. The large dust lane seen in the HST image to the South East of NGC838 is clearly visible. A color version of this Figure is available in the online version of the article.}\label{fig:hb-ha}
\end{figure}

To estimate the extragalactic reddening along a given line of sight, we compute the total extinction in the V band A$_V$ (in magnitudes) using the H$\alpha$ to H$\beta$ flux ratio. The resulting map is shown in Figure~\ref{fig:hb-ha}, where only the 558 spaxels with S/N(H$\alpha$,H$\beta$)$\geq$3 are displayed.

NGC838 contains several large dust lanes clearly visible in the HST image shown in Figure~\ref{fig:hst}. In the reddening map, the impact of the most prominent dust lane, located to the South East of the galaxy center, can be readily identified as the region with A$_V$$\geq$ 2.5 mag. The least obscured line-of sights are located South of this area, with A$_V\leq$1 mag. To the North West, the extinction is rather uniform with A$_{V}\cong$2 mag. This uniform extinction region contrasts with the thick dust lanes visible to the North West of NGC838 in Figure~\ref{fig:hst}. The lack of well defined dust lanes in this area of the reddening map indicates that the ionized gas is located above the central region of NGC838, consistent with an outflow scenario.

The quoted A$_V$ values in Figure~\ref{fig:hb-ha} assume a case B recombination and associated intrinsinc H$\alpha$ to H$\beta$ flux ratio of 2.85, typical of H{\footnotesize II} regions. This is a legitimate value because NGC838 is not hosting any AGN (which have been shown can induce larger intrinsic H$\alpha$ to H$\beta$ ratios, see \cite{Kewley06}). A$_V$ values have been computed using the \cite{Fischera05} extinction curve with R$_V^A$=4.5, which provides self-consistent SFR estimates between UV, H$\alpha$ and [OII] luminosities \citep[][]{Wijesinghe11}. This theoretical extinction curve assumes the presence of a turbulent foreground dust screen, and (with R$_V^A$=4.5) is a good match to the empirical extinction curve for starburst galaxies of \cite{Calzetti01}. The exact equation of the extinction curve is derived in the Appendix, see Eq.~\ref{eq:final}. 

\begin{figure*}[htb!]
\centerline{\includegraphics[scale=0.5]{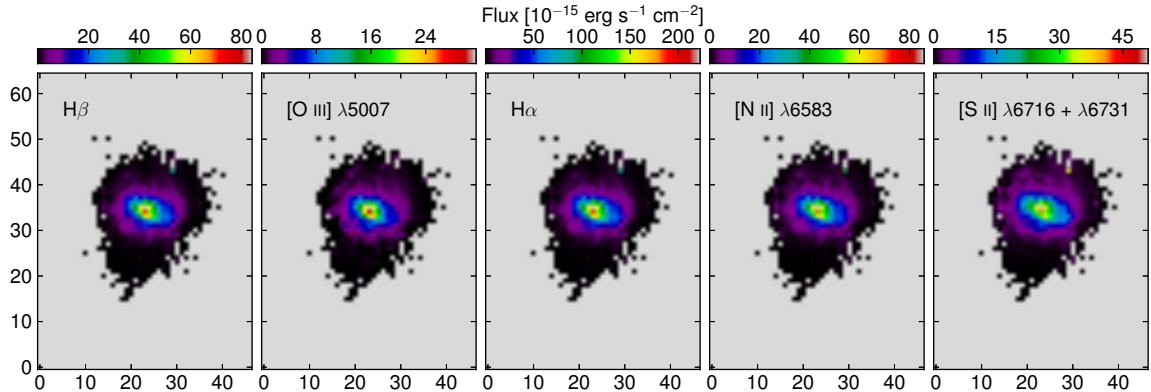}}
\caption{Integrated flux maps for the H$\beta$, [O {\scriptsize III}]  $\lambda$5007, H$\alpha$, [N {\scriptsize II}] $\lambda$6583 and  [S {\scriptsize II}] $\lambda$6716 + $\lambda$6731 lines after correcting for galactic and extragalactic extinction for spaxels with S/N(H$\beta$,H$\alpha$)$\geq$3. Our original flux detection limit is of the order of $\sim$1$\times10^{-17}$ erg s$^{-1}$ cm$^{-2}$. A color version of this Figure is available in the online version of the article.}\label{fig:linemaps}
\end{figure*}

Using Eq.~\ref{eq:corr}, we corrected each spectrum for extragalactic reddening using the associated H$\alpha$ to H$\beta$ flux ratio. The resulting emission line maps for the H$\alpha$, H$\beta$, [O {\footnotesize III}]  $\lambda$5007, [N {\footnotesize II}] $\lambda$6583 and [S {\footnotesize II}] $\lambda$6716 + $\lambda$6731 are shown in Figure~\ref{fig:linemaps}. Only spaxels with S/N(H$\alpha$,H$\beta$)$\geq$3 for which an accurate reddening correction can be computed are shown. The corresponding flux density lower detection limit of emission lines in our observations is $\sim$1$\times10^{-17}$ erg s$^{-1}$ cm$^{-2}$ \AA$^{-1}$ per spaxel. For a resolution of R=7000, this line peak limit is equivalent to a minimal detectable flux (per spaxel) from an unresolved emission line (with $\sigma\leq$25 km s$^{-1}$) of $\sim$1.3$\times10^{-17}$ erg s$^{-1}$ cm$^{-2}$ at 6000 \AA.

The total H$\alpha$ flux of NGC838 is $\log$(F$_{H\alpha})$ = 42.3 erg s$^{-1}$ cm$^{-2}$ after correcting for extragalactic reddening, and $\log$(F$_{H\alpha})$ = 41.5 erg s$^{-1}$ cm$^{-2}$ before. This value is in perfect agreement with \cite{Meurer06} who (correcting only for galactic extinction) obtain a total flux of $\log$(F$_{H\alpha})$ = 41.6 erg s$^{-1}$ cm$^{-2}$ for NGC838.

\section{Results}\label{sec:results}

\subsection{The underlying stellar population of NGC838}

\begin{figure*}[htb!]
\centerline{\includegraphics[scale=0.4]{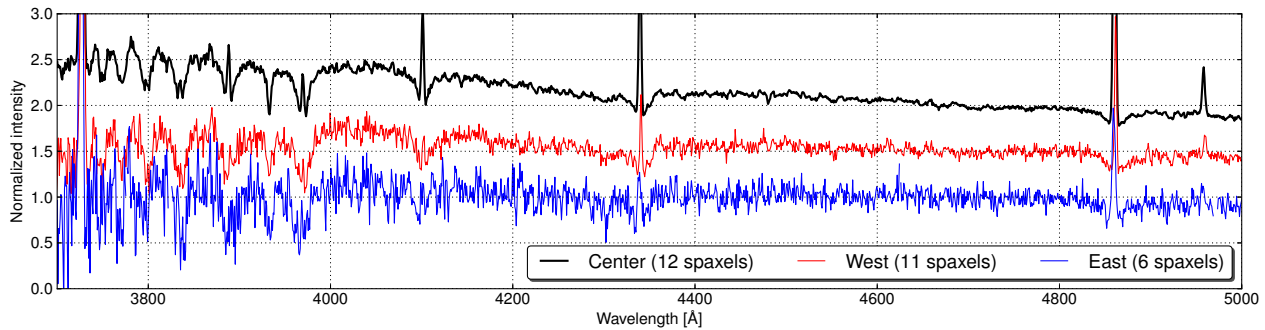}}
\caption{Spectra extracted from three apertures centered at the positions (23;34 - upper spectrum), (33,32 - middle spectrum) and (15,38 - lower spectrum) and 4" in diameter. Each spectrum has been normalized at 4700\AA. The upper and middle spectra have been offset by +1 and +0.5 respectively. Only 11 and 6 spectra with S/N(H$\beta$,H$\alpha$)$\geq$3 where summed in the West and East apertures. A color version of this Figure is available in the online version.}\label{fig:A0spec}
\end{figure*}

In Figure~\ref{fig:A0spec}, we show three spectra extracted from three apertures located in the center (top, thick line), to the West (middle line) and to the East (bottom line) of NGC838. The apertures are centred at the spaxel coordinates (23,34), (33,32) and (15,38) and are 4" in diameter. Each spectrum has been normalized at 4700{\AA}, and for clarity, the top and middle spectra have been offset by +1 and +0.5 respectively. The wavelength scale is in the rest frame of the galaxy, assuming v$_{\text{NGC838}}$=3851 km s$^{-1}$ (see Table~\ref{table:hcg16}).

Despite a lower S/N in the lower two spectra, a clear signature of a young A-type stellar population is visible in all three spectra. The strength of the absorption features associated with this young \citep[$\sim$500 Myr post-starburst, e.g.][]{Boehm92,Stahler05} stellar population is not diminishing away from the galaxy center. The spectrum extracted from the central region of NGC838 has a stronger blue continuum compared to the East and West ones. This signature of O/B-type star population is consistent with a very recent/ongoing starburst in the central region of NGC838. These young and hot stars are visible in the HST image of NGC838 as a central patch of blue emission. These features suggest that NGC838 has been subject to an episode of star formation throughout the whole extend of the galaxy some $\sim$500 Myr ago, and that the activity has now settled down in the central regions of the galaxy.

\subsection{Velocity structure of the ionized gas}\label{sec:velstruct}

\begin{figure*}[htb!]
\centerline{\includegraphics[scale=0.5]{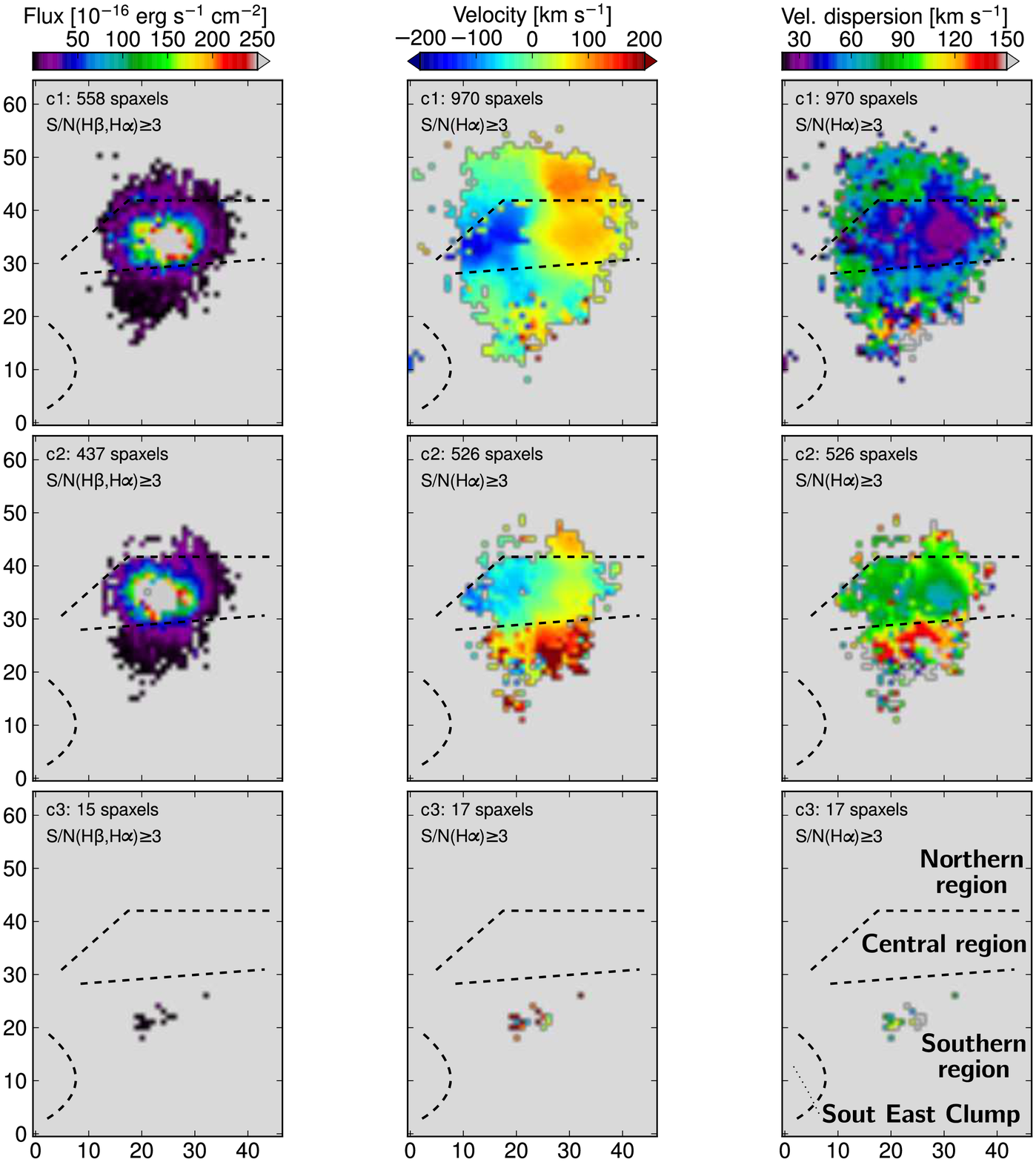}}
\caption{H$\alpha$ flux (left), velocity (center) and velocity dispersion (right) maps of NGC838. Each row corresponds to a different gaussian component of the emission line, with maps labelled accordingly. Dashed lines mark the divisions between the four regions (Northern, Central, Southern and the South East Clump), labelled accordingly in the $c3$ velocity dispersion map (bottom right). The number of pixels in the maps and the associated S/N cut-offs are written in each panel. A color version of this Figure is available in the online version of the article.}\label{fig:vsigmaps}
\end{figure*}

The H$\alpha$ flux, velocity and velocity dispersion maps for the $c1$ (lowest velocity dispersion), $c2$ and $c3$ (highest velocity dispersion) groups of gaussian components fitted to the emission lines are shown in Figure~\ref{fig:vsigmaps}. Velocities are in the NGC838 rest frame. A heliocentric velocity correction of $\sim$2 km s$^{-1}$ was taken into account. Velocity dispersion have been corrected for instrumental resolution ($\sim$1{\AA} at 6500{\AA}).

Our extragalactic reddening correction relies on the H$\beta$ to H$\alpha$ flux ratio, and only spaxels with S/N(H$\beta$,H$\alpha$)$\geq$5 can be corrected with enough accuracy to perform subsequent line ratio analysis (see Sec.~\ref{sec:lineratios}). Consequently, our flux intensity maps contain less spaxels than the corresponding velocity and velocity dispersion maps, for which the cutoff criteria is S/N(H$\alpha$)$\geq$3. 

The $c1$ and $c2$ velocity maps reveal a complex mixture of red- and blueshifted regions. A marked discontinuity in the $c1$ flux map is visible at the transition region between 1 and 2 components. This discontinuity suggest that multiple components in the emission lines are present further away from the galaxy core than detected in our data, but cannot be resolved clearly due to the decreasing S/N beyond $\sim$10 spaxels from the center. The structure of the $c1$ and $c2$ velocity and velocity dispersion maps suggest a subdivision into four regions, delineated by dashed lines in Figure~\ref{fig:vsigmaps}. We refer to them as the \emph{South East Clump} and the \emph{Northern, Central and Southern regions}. 

\paragraph{ The South East Clump}\label{sec:clump}
This region only comprises 5 spaxels on the eastern edge of our mosaic. Its H$\alpha$ line is well fitted by one gaussian component only. The associated velocity dispersion is low, of the order of 20-30 km s$^{-1}$. A close inspection of the H$\alpha$ SINGG image of NGC838 (see Figure~\ref{fig:singg}) reveals some possible features in this region, but the noise in the image makes these hard to identify with certainty. The clump is blueshifted, with a mean radial velocity equal to -122$^{\pm6}$ km s$^{-1}$. 

\paragraph{The Central region}\label{sec:cz}
This area encompasses the center of NGC838. In the $c1$ component, it has a low velocity dispersion ($\sim$25-50 km s$^{-1}$) and a well defined signature of organised rotation. In the $c2$ component, a weaker rotation signature can also be identified, with a higher velocity dispersion of the order of 90 km s$^{-1}$.

\paragraph{The Northern region}\label{sec:nz}
Most of this region is fitted with only one component, with the exception of a small region to the West. The associated velocity dispersion is of the order of 90 km s$^{-1}$, and a rotation signature can be identified in the $c1$ velocity map. 

\paragraph{The Southern region}\label{sec:sz}
This region contains very high velocity dispersion values in both the $c1$ and $c2$ maps, ranging from 90-150 km s$^{-1}$ in the southernmost parts. This region does not display any sign of organised rotation in either the $c1$ or $c2$ maps, but is strongly redshifted with velocities $\sim$400 km s$^{-1}$, seen in the $c2$ velocity map. All spaxels fitted with three components are located in this region.

In the next two Sections, we examine in detail the kinematics of the Central, Northern and Southern regions.

\subsubsection{The rotation signature of the Northern and Central regions}\label{sec:nszones}

\begin{figure}[htb!]
\centerline{\includegraphics[scale=0.33]{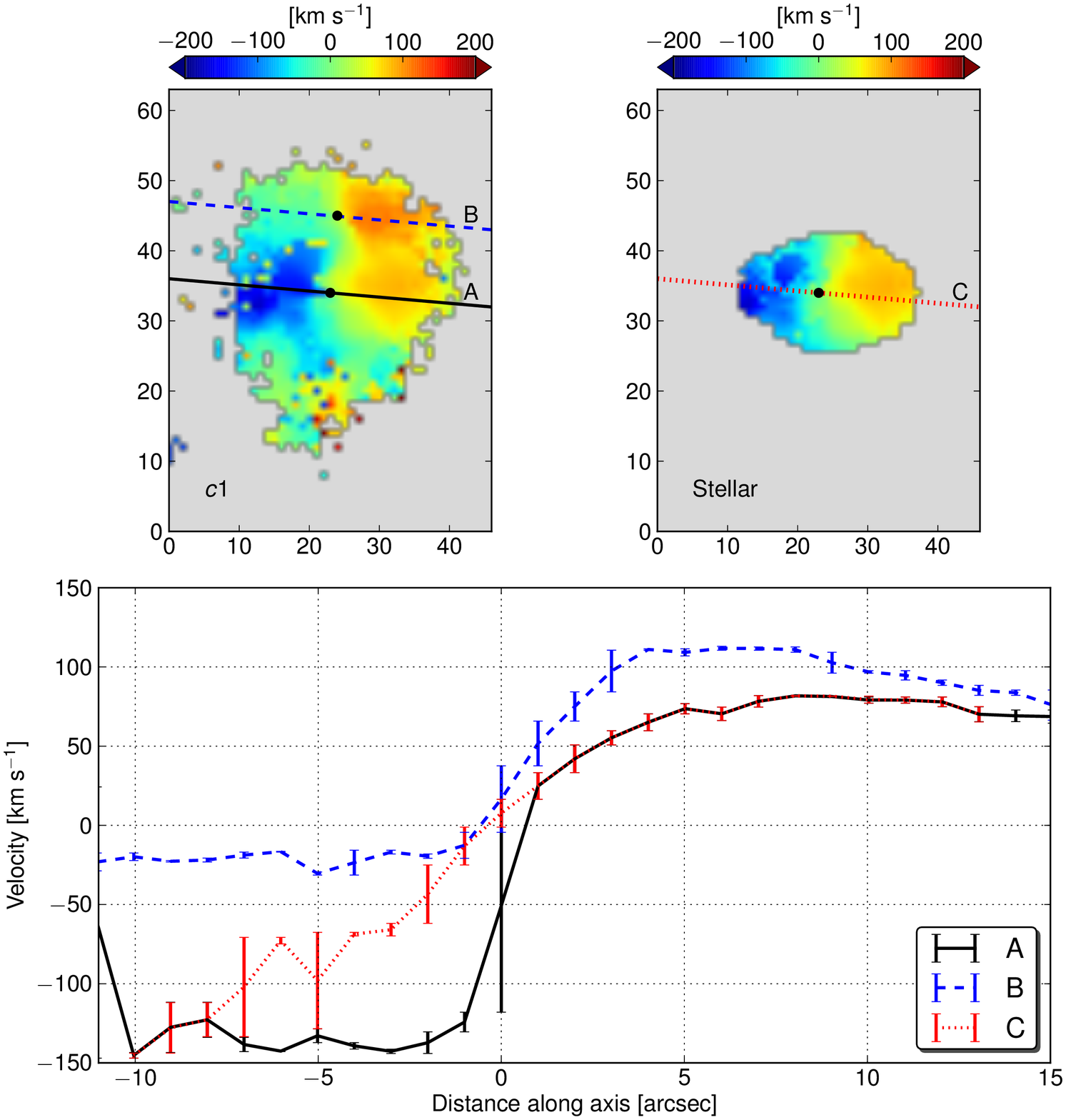}}
\caption{Top : $c1$ velocity map of H$\alpha$ (left) and stellar velocity map (right). The A (full), B (dashed) and C (dotted) lines are the axes along which the radial velocity curves are extracted. Bottom : extracted radial velocity profiles. A color version of this Figure is available in the online version of the article. }\label{fig:radial}
\end{figure}

To best reveal the rotation signature of the Central and Northern region, we extract their associated rotation curve, and show them in the lower panel of Figure~\ref{fig:radial}. Each rotation curve is extracted along the relevant axis in the two top panels which show the H$\alpha$ $c1$ velocity map on the left and the stellar velocity map on the right. The stellar velocities are obtained from the stellar template fitting routine within UHSPECFIT. In each case, velocity values are extracted in a 3 pixels band around each axis and averaged at every position. The associated 1-$\sigma$ error is shown for every position. In the velocity maps (upper panels), the black points denote the location of the reference \emph{zero} position on each axis, with the distance increasing positively towards the West (right).

For the Central Region, we set the reference \emph{zero} location at the spaxel coordinate (23;34). This corresponds to the H$\alpha$ intensity peak location. The axis is inclined at 95$^{\circ}$ West of North, our estimate of the P.A. of NGC838  using our flux maps (see Figure~\ref{fig:linemaps}). This value is consistent with the NED \citep[][]{Helou91} values for the P.A. of NGC838 (81 degree in the r-band, 95 degrees in the Ks band). The (projected) rotation velocities range from +80 km s$^{-1}$ to -140 km s$^{-1}$, spanning a total range of 220 km s$^{-1}$. The central transition region between the red- and blueshifted side is very sharp, and the rotation curve flattens out rapidly further out. This well defined rotation signature most certainly arises from low-dispersion gas in orbit within the plane of the galaxy. To confirm this interpretation, we extract a velocity curve along the same axis using stellar velocities derived from the stellar velocity map (shown in Figure~\ref{fig:radial}, upper right panel). This velocity map appears smoother than the H$\alpha$ velocity field in the central regions. The structure visible in the stellar velocity map to the East of the galaxy center is most likely a consequence of the small S/N in this area, which is also subject to high extinction values (see Figure~\ref{fig:hb-ha}). The stellar rotation curve ranges from $\sim$-100 km s$^{-1}$ to +80 km s$^{-1}$ and is consistent with the gas rotation curve. Both rotation curves have an overall offset of -30 km s$^{-1}$ with respect to the rest frame velocity of NGC838.

The rotation curve in the Northern region is extracted along an axis inclined at 95$^{\circ}$ West from North and centered at the spaxel coordinate (23,46). Its orientation is taken to be that of the axis along which the Central region rotation curve is extracted. The (projected) rotation velocity curve goes from $\sim$-30 km s$^{-1}$ to +110 km s$^{-1}$, spanning a total range of 140 km s$^{-1}$. The left-hand side of the curve is flat. The right-hand side, after peaking at 140 km s$^{-1}$, is decreasing towards a velocity of 100 km s$^{-1}$. The Northern region rotation curve has an overall redshift of +40 km s$^{-1}$ with respect to the adopted rest frame velocity of NGC838, or +70 km s$^{-1}$ compared to the stellar rotation curve. Previous studies of NGC838 which failed to distinguish between these multiple components led to erroneous estimates of its radial velocity. From our observations, it is clear that at least 2 distinct gas populations are present with very distinct kinematic properties. From our measurement of the rotation curve of both the H$\alpha$ emitting gas and the stellar population, we find that the true radial velocity of NGC838 is most likely 30 km s$^{-1}$ slower ($\sim$3820 km s$^{-1}$) than the current value found in the literature \citep[$\sim$3850 km s$^{-1}$ in SDSS DR9, see][]{Ahn12}, highlighting the power of wide integral field spectroscopy for velocity field characterization. This radial velocity change does not challenge the membership of NGC838 to HCG16. We will discuss a possible explanation for the 70 km s$^{-1}$ velocity offset between the Central and Northern regions in Sec.~\ref{sec:finalpicture}

\subsubsection{The redshifted Southern region}

\begin{figure}[htb!]
\centerline{\includegraphics[scale=0.38]{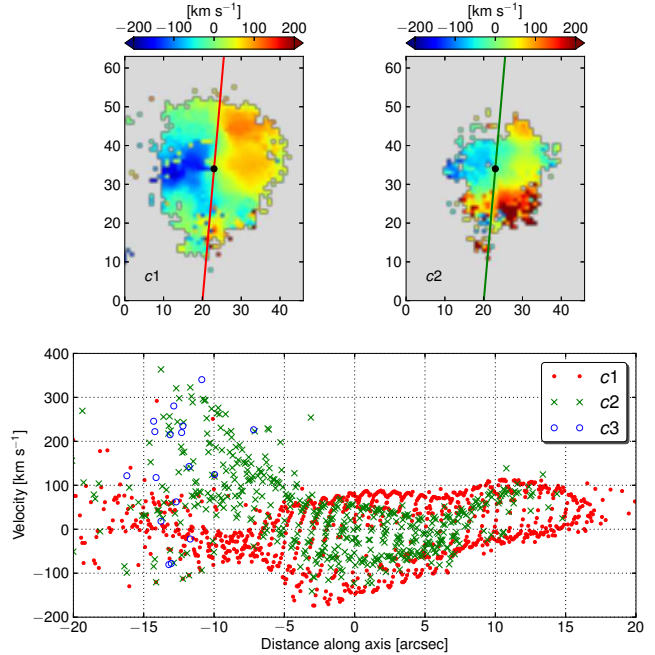}}
\caption{PV diagram (lower panel) for the $c1$ (lowest velocity dispersion, black dots, in red in the online version), $c2$ (grey crosses, in green in the online version) and $c3$ (highest velocity dispersion, black circles, in blue in the online version) velocity maps of NCG838. The projection axis is shown in the two velocity maps (upper panels), rotated 5$^{\circ}$ West from North and with the reference \emph{zero} location at the position (23,34).}\label{fig:pv}
\end{figure}

To best reveal the complex velocity structure of the Southern region, we construct a position-velocity (PV) diagram of the $c1$, $c2$ and $c3$ velocity maps, shown in Figure~\ref{fig:pv}. PV diagrams are a powerful tools which can simplify the visualisation of IFS 3-dimensional data by \emph{removing} a spatial dimension. The concept is as follows. Every spaxel is projected onto a given axis, and assigned the corresponding distance from the reference location on this axis. This distance is then shown against the spaxel velocity in the PV diagram, which (with a carefully chosen axis) can reveal complex  structures in the data cube.

The projection axis chosen to construct our PV diagram is shown in the two velocity maps in Figure~\ref{fig:pv} (upper panels). It is perpendicular to the radial velocity axis in Figure~\ref{fig:radial}, centered at the spaxel coordinate (23,34) and rotated 5$^{\circ}$ West of North. In the PV diagram (lower panel), red points correspond to the $c1$ pixels (lowest velocity dispersion), green crosses to the $c2$ pixels, and blue circles to the $c3$ pixels (highest velocity dispersion). 

The high redshift values present in the Southern region result in the large spread of $c2$ spaxels to the left of the PV diagram. The most redshifted spaxels have velocities slightly below 400 km s$^{-1}$. The distribution of both the $c1$ and $c2$ spaxels does not follow any specific pattern for distances smaller than -5 pixels along the projection axis. This lack of structure is in strong contrast to the Central (-5 pixels to +7 pixels along the projection axis) and Northern (+7 pixels to +18 pixels along the projection axis) regions, which are organised in well defined structures. The barred tubular structures are the (expected) result of the rotation of these two regions. The size of the envelope is equal to twice the (projected) rotation velocity of the gas. The offset between the Central and Northern region mean velocities of 70 km s$^{-1}$ is reflected in the discontinuity visible at around +7 pixels along the projection axis in the red dot distribution. In addition, the smaller rotation velocity of the Northern region results in a cylinder of red points thiner beyond a distance of +7 pixels from the reference point. In the Central region, the green crosses trace another small barred tubular structure, associated with the weak rotation of the $c2$ component in this region seen previously in Figure~\ref{fig:vsigmaps}.

\subsection{Velocity structure of the neutral gas}\label{sec:ned}

Galactic winds can be detected and studied via absorption lines, for example using the Na I $\lambda\lambda$5890,5896 doublet \citep[the Na {\footnotesize D} lines, see, e.g. ][]{Heckman00,Rupke02}. Depending on the galaxy orientation, Na {\footnotesize D} lines can probe either neutral gas within the disk (in edge-on systems) or material entrained within the wind \citep[in face-on systems, see][]{Chen10}. The NED inclination for NGC838 based on Ks band observations from the 2MASS survey is 46 degrees. This low inclination angle suggest that in NGC838, the Na {\footnotesize D} absorption lines will primarily have their origin in material within the disk.  

\begin{figure}[htb!]
\centerline{\includegraphics[scale=0.6]{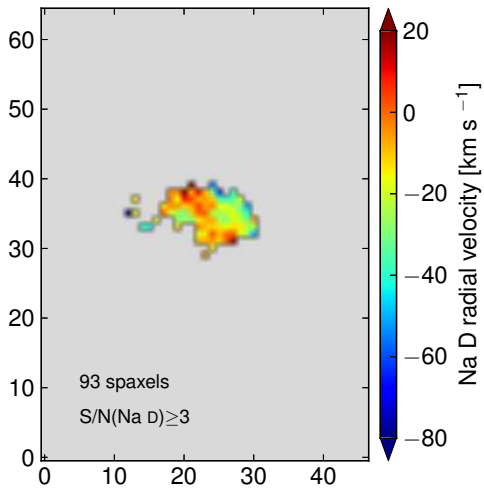}}
\caption{Velocity map of the Na {\scriptsize D} doublet. The velocities are given in the rest frame of NGC838. The colorbar has been chosen so that it matches our measured radial velocity of NGC838. Black and white regions (red and blue in the online version) correspond to red- and blueshifted regions with respect to the mean redshift of the stellar rotation curve of NGC838 (see Sec.~\ref{sec:nszones}).}\label{fig:nad}
\end{figure}

We fit the Na {\footnotesize D} doublet for all spaxels with S/N(Na {\footnotesize D})$\geq$5 (see Figure~\ref{fig:fit}). The resulting velocity map is shown in Figure~\ref{fig:nad}. The Na {\footnotesize D} lines did not show any asymmetry or multiple line profile in any spaxel. They were always best fitted by single gaussian. For practical reasons, the Na {\footnotesize D} fits were performed separately from that of the emission lines using a dedicated Python routine. The quoted velocities are in the rest frame of NGC838, assuming v$_{\text{NGC838}}$=3851 km s$^{-1}$ its radial velocity found in the literature. However, the colorbar is offset by 30 km s$^{-1}$ (the offset we identified in Sec.~\ref{sec:nszones}) to allow for a better identification of inflows (red colours) and outflows (blue colours) with respect to the disk of NGC838.

The velocity map reveals two distinct kinematic regions for the neutral gas with velocities of the order of +10 to -30 km s$^{-1}$. We do not detect any fast outflow signature, consistent with the inclination of NGC838. To the East, a band of gas is redshifted by 20-30 km s$^{-1}$ with respect to the disk mean velocity. To the West, the gas velocity is of the order of the mean stellar velocity of NGC838. A comparison with the HST image of NGC 838 and our extinction map in Figure~\ref{fig:hb-ha} suggests that the East band may be the signature of gas inflow towards the galaxy center. This inflow is spatially coherent with the large dust lane to the East of the galaxy center. This dust lane may be shielding the neutral gas from the main outflow. Further away from the dust lane (towards the North West), hints of a blueshifted neutral gas component suggest that the wind bubble's expansion is pushing the neutral gas away as it expands. This interpretation is subject to caution as the Na {\footnotesize D} velocity map is small. Higher S/N observations are required to confirm this scenario.

\subsection{Line ratio maps and diagnostics}\label{sec:lineratios}
\begin{figure}[htb!]
\centerline{\includegraphics[scale=0.5]{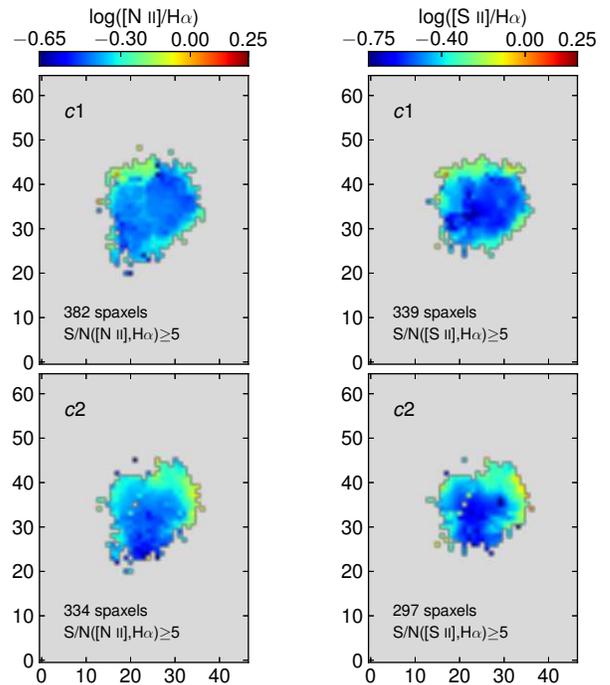}}
\caption{Line ratio maps for [N {\scriptsize II}]/H$\alpha$ (left) and [S {\scriptsize II}]/H$\alpha$ (right). The upper panels correspond to the $c1$ component, and the lower panel to the $c2$ component. A color version of this Figure is available in the online version of this article.}\label{fig:haratio}
\end{figure}

Specific line ratios can help understand the dominant excitation mechanism of the outflow in NGC838 \citep[][]{Baldwin81,Veilleux87, Veilleux02,Matsubayashi09}. We use the [N {\footnotesize II}]/H$\alpha$, [S {\footnotesize II}]/H$\alpha$, [O {\footnotesize I}]/H$\alpha$ and [O {\footnotesize III}]/H$\beta$ ratios. These ratios are not strongly dependant on reddening corrections, a significant advantage in the case of NGC838 which is for the most part severely affected by intrinsic dust absorption (see Figure~\ref{fig:hb-ha}). We show the line ratio maps for [N {\footnotesize II}]/H$\alpha$ and [S {\footnotesize II}]/H$\alpha$ in Figure~\ref{fig:haratio}. All three lines involved in these ratios are located in the red part of our spectra (high resolution), and the ratios for individual $c1$, $c2$ and $c3$ components can be computed. Because the $c3$ map is very scarce and does not contain critical information, we only show the line ratios for the $c1$ (upper panels) and $c2$ (lower panels) components. Only spaxels with S/N([N {\footnotesize II}],H$\alpha$)$\geq$5 and S/N([S {\footnotesize II}],H$\alpha$)$\geq$5 are shown. 

All four maps display the same general structure. Low line ratio values are located towards the center, and higher values towards the edges. For the $c1$ component (lowest velocity dispersion), the map edges are thin ($\sim$2-5") with high line ratios. For the $c2$ maps, the high ratio value edges are thicker ($\sim$5-10"), but not located all around the maps. A tongue of low line ratios is extending towards the South East all the way to the edge of the maps in both the [N {\footnotesize II}]/H$\alpha$ and [S {\footnotesize II}]/H$\alpha$ $c2$ maps. 

\begin{figure}[htb!]
\centerline{\includegraphics[scale=0.5]{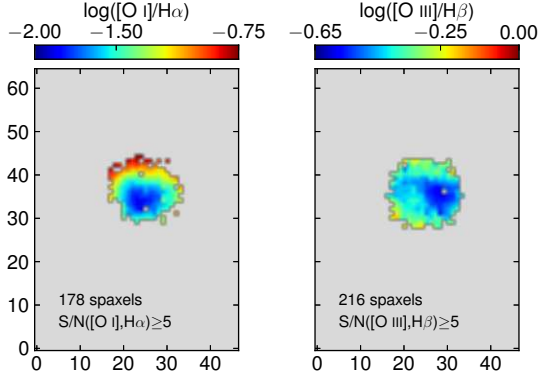}}
\caption{Line ratio maps for [O {\scriptsize I}]/H$\alpha$ (left) and [O {\scriptsize III}]/H$\beta$ (right). A color version of this Figure is available in the online version of the article.}\label{fig:hbratio}
\end{figure}

In Figure~\ref{fig:hbratio}, we show the [O {\footnotesize I}]/H$\alpha$ and [O {\footnotesize III}]/H$\beta$ line ratio maps. Because (1) S/N([O {\footnotesize I}])$\cong$5 and (2) the [O {\footnotesize III}] and H$\beta$ lines are located in the blue part of our spectra (low resolution), only the integrated, total flux ratios are computed.

The [O {\footnotesize I}]/H$\alpha$ map shows a radial gradient with the lowest ratios centered on NGC838. We note the lack of any strong [O {\footnotesize I}] emission in the Southern region. For [O {\footnotesize III}]/H$\beta$, the lowest ratios are located $\sim$5" West of the Galaxy center. A core of small ratio values ($\log$([O {\footnotesize III}]/H$\beta$)$\cong$-0.6) is surrounded by a thick envelope of higher values ($\log$([O {\footnotesize III}]/H$\beta$)$\cong$-0.3).

\begin{figure}[htb!]
\centerline{\includegraphics[scale=0.58]{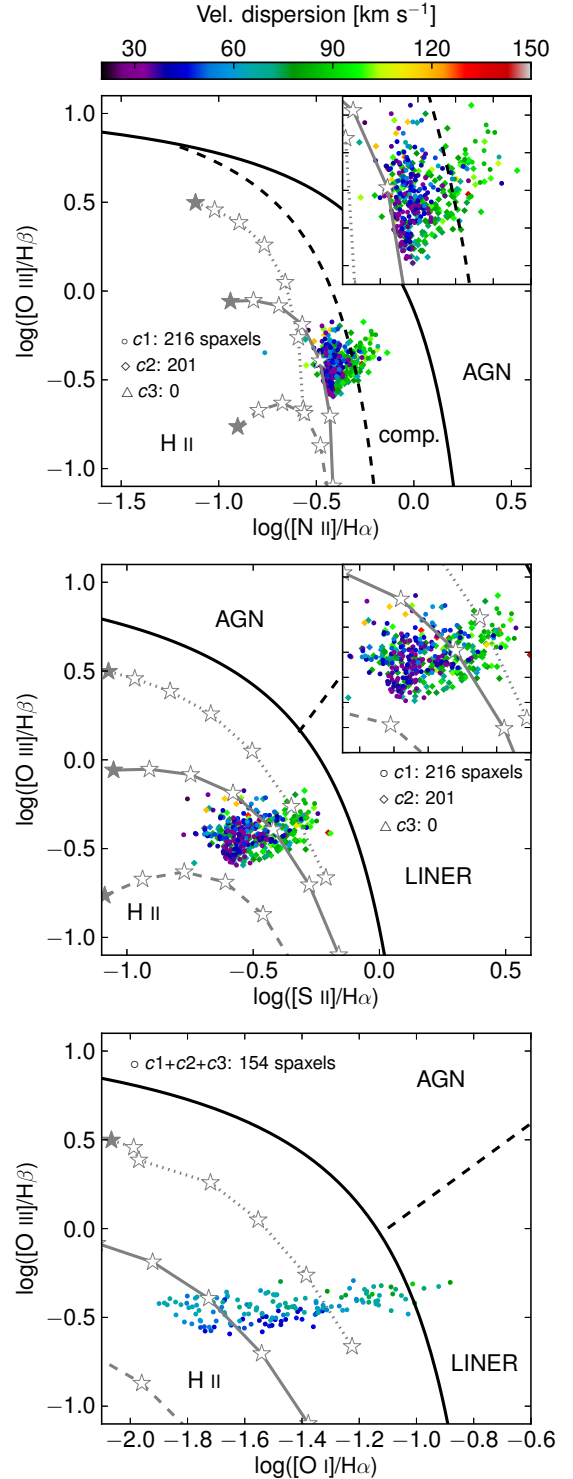}}
\caption{Standard line ratios diagnostic diagrams for NGC838. In the upper two panels, circles correspond to $c1$ (low $\sigma$) line ratios, diamonds to the $c2$ line ratios, and triangles to the $c3$ line ratios. In the lower panel, only the total flux line ratios are used, resulting in only one data point per spaxel. The H{\scriptsize II}, AGN and LINER region markers are from \cite{Kewley06}. Points are colored according to their associated velocity dispersion. Stars are photoionization models for solar (dotted line), twice solar (full line) and three times solar (dashed line) metallicity and a varying ionization factor Q. Filled stars correspond to Q=8. A color version of this Figure is available in the online version of this article.}\label{fig:bpt}
\end{figure}

We show the corresponding line ratio diagnostic diagrams in Figure~\ref{fig:bpt}. Each data point corresponds to one spaxel in our data cube. Because any given spaxel has up to three different associated [N {\footnotesize II}]/H$\alpha$ and [S {\footnotesize II}]/H$\alpha$, but only one value of [O {\footnotesize III}]/H$\beta$, only the top two diagrams have three data sets visible: $c1$ (lowest velocity dispersion, circles), $c2$ (diamonds) and $c3$ (highest velocity dispersion, triangle). The same [O {\footnotesize III}]/H$\beta$ total flux ratio is associated with all three (different) $c1$, $c2$ and $c3$ [N {\footnotesize II}]/H$\alpha$ and [S {\footnotesize II}]/H$\alpha$ ratios. In fact, no $c3$ point is shown because none of these spaxels have S/N([O {\footnotesize III}],$H\beta$)$\geq$5. The different excitation regions defined in \cite{Kewley06} are shown. Each data point is coloured according to its velocity dispersion. The colorbar extent and stretch is identical to Figure~\ref{fig:vsigmaps}. For the lower diagnostic diagram, the velocity dispersion of each spaxel is taken as the H$\alpha$ flux weighted mean velocity dispersion of the spaxels $c1$, $c2$ and $c3$ associated components. In the upper and central panel, a zoom on the region of interest is shown in a sub panel of the main diagrams. Photoionization models computed with the Starburst99 \citep[][]{Leitherer99} and MAPPINGS III \citep[][]{Sutherland93} codes for ionization parameters Q$\in$[6.50,6.75,7.00,7.25,7.50,7.75,8.00] are shown with stars symbols. The dotted line corresponds to solar metallicity, the full line to twice solar, and the dashed line to three times solar.

All three diagnostic diagrams indicate that photoionization is the main excitation mechanism for the gas in the center of NGC838 (see Figure~\ref{fig:haratio}~\&~\ref{fig:hbratio}). In the upper and central panels, a clear dichotomy exists between low- and high-velocity dispersion spaxels. Two branches are visible in the upper panel - a low velocity dispersion branch ($\sim$30-50 km s$^{-1}$) parallel to the H{\footnotesize II} region delimiting line, and a high velocity dispersion branch ($\sim$30-50 km s$^{-1}$) extending towards and entering the composite region. We note that the former of these two branches contain in fact both low and high velocity dispersion points, while the second one only contains a high velocity dispersion component. These two branches can be interpreted as the respective signatures of (1) pure photoionization of gas within the disk of the galaxy and at the base of the wind and (2) mixing between photoionization and slow shocks along the wind axis. The photoionization branch is consistent with the twice solar metallicity photoionization model and an ionization parameter Q=7.0. In their study of galaxy-wide shocks in late-stage mergers, \cite{Rich11} computed theoretical line ratios for varying mixing values between pure photoionization and pure shock excitation. Comparing with Figure 10 in their paper, the composite line ratio branch that we detect in NGC838 is consistent with up to 20\% shock fraction.

In the lower panel, log([O {\footnotesize I}]/H$\alpha$) is extending from -2.0 to -0.9, a wide range compared to the stretch of any other line ratio. Four data points also extend into the LINER region. These extreme ratio values are located away from the galaxy center (see Figure~\ref{fig:hbratio}). The increased [O {\footnotesize I}] line strength, reproducing to some extent a LINER signature, is a consequence of the geometry of the ionization bubble in NGC838. Towards the edge of the bubble, the geometrical depth along the ionization front is larger than towards the center. Overall, our line ratios are consistent with previous measurements from \cite{Carvalho99} for NGC838. 

\section{Discussion}\label{sec:discussion}

\subsection{The global picture}\label{sec:finalpicture}

\begin{figure}[htb!]
\centerline{\includegraphics[scale=0.3]{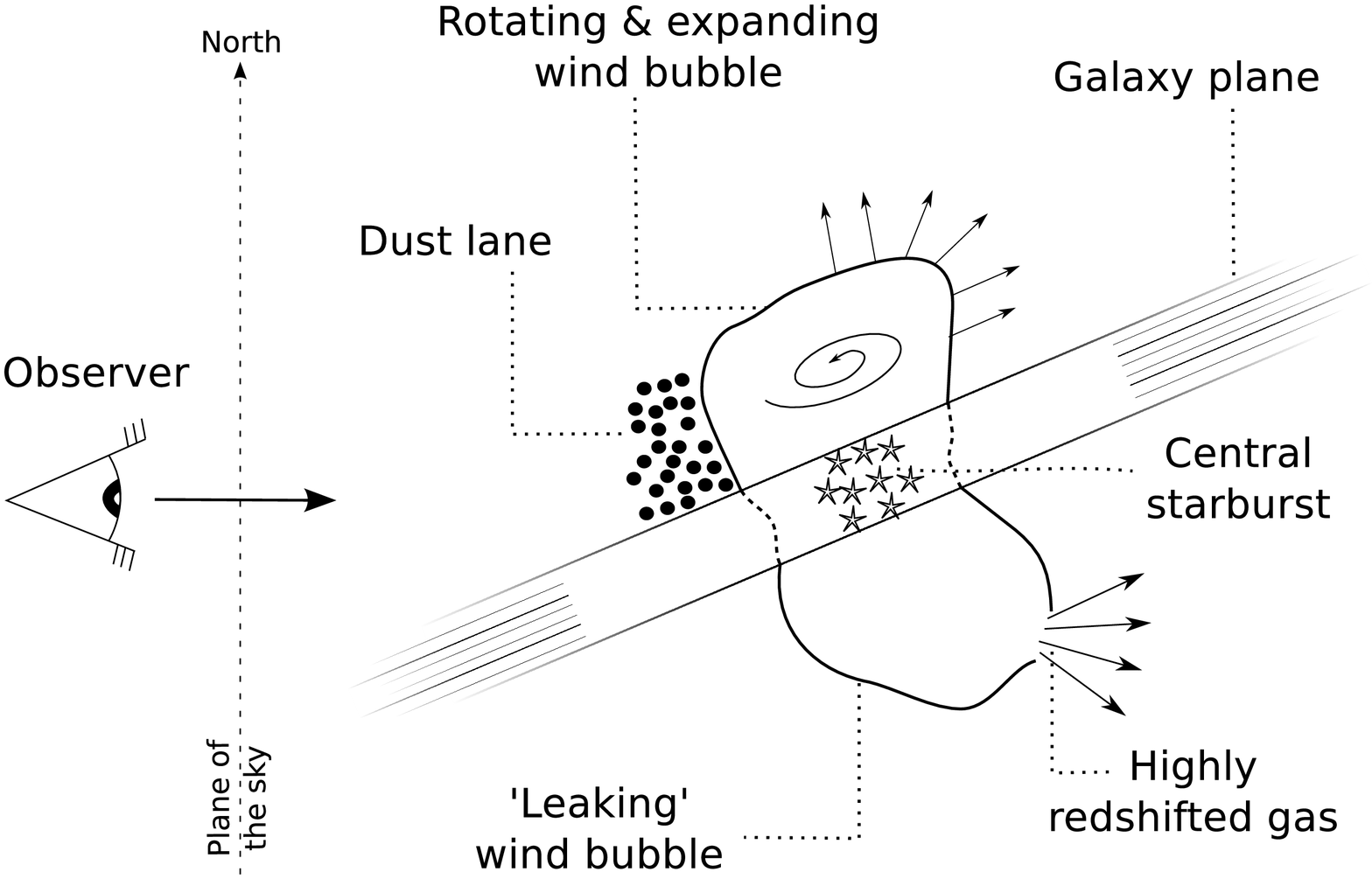}}
\caption{Global schematic of the galactic wind in NGC838, with an expanding and rotating Northern wind bubble, a Southern wind bubble just entering its blowout phase, both of which are powered by a central starburst (not to scale).}\label{fig:schema}
\end{figure}

Previous, low-resolution observations of NGC838 suggested that it has a double nucleus with a 2" separation and a complex velocity structure. These observations were interpreted as the signature of an ongoing merger \citep[][]{Mendes98}. We believe that Figure~\ref{fig:hst} and our WiFeS observations argue against this merger scenario, even if longer range interactions with neighbouring galaxies cannot be ruled out. We interpret the complex velocity structure detected previously as the signature of a gaseous outflow powered by a nuclear starburst, in a similar fashion to its neighbour NGC839 \citep[][]{Rich10} or the archetypical M82 \citep[][]{Shopbell98}. The nuclear starburst is seen as an extended population of young, hot, blue stars in the HST image. We present in Figure~\ref{fig:schema} a toy model for the overall system which illustrates the different elements of our observations.

From the lack of clearly defined features in the extinction map (aside from the dominant dust lane to the South West, see Sec.~\ref{sec:red}), we deduce that the gas responsible for the Northern region emission seen in our observation is located above the galaxy plane, consistent with its outflow nature. Gas associated with this Northern region outflow has a well defined rotation (see Sec.~\ref{sec:nszones}). The rotation is similar to that of the stellar component of NGC838, yet with a peak velocity 40 km s$^{-1}$ slower and in a rest frame offset by 70 km s$^{-1}$. A galactic wind rotating slightly slower than its host galaxy is consistent with previous observations of other galactic winds \citep[e.g. in M82, see][]{Shopbell98,Westmoquette09}, as well as with theoretical models \citep[][]{Chatt12}. As the wind expands, conservation of angular momentum requires this to be the case. 

The HST image of NGC838 and its dust lanes makes the orientation of the galaxy un-ambiguous, placing the Southern side of the galaxy in front (see Figure~\ref{fig:schema}). In that geometry, one would expect to see a blueshifted wind blowing towards the Earth in the Northern region. The 70 km s$^{-1}$ mean redshift of the outflow in the Northern region with respect to stellar disk mean velocity is therefore puzzling. We propose the following explanation for the wind mean redshift in this region. Thick dust lanes at the base of the wind act as barriers, and deflect the outflow toward the back of the galaxy. This scenario is motivated by the very thick dust lane to the South West of NGC838. It is also consistent with the velocity of the neutral gas in NGC838 (see Figure~\ref{fig:nad}). Neutral gas is falling-in towards the nucleus behind the dust lane, but transitions to a blueshifted velocity away from the dust lane (towards the North East), where the wind bubble expands more easily. \cite{Cooper08} have shown in their simulations that high gas (and dust) densities can significantly reduce the expansion of a wind bubble along the major axis of a galaxy. 

The Southern region contains highly redshifted, non-rotating gas, which is consistent with a \emph{leaking wind bubble} or \emph{blow-out}. In this scenario, the wind bubble, powered by the nuclear starburst, has bursted out of the surrounding H{\footnotesize I} envelope into a lower density environment \citep[e.g.][]{Veilleux05}. The high pressure inside the wind bubble is driving the rapid escape of the hot gas it contains and suppresses any existing wind rotation. We detected hints of multiple ($\geq3$), distinct, velocity component in the Southern region, the possible signature of individual clumps being accelerated in the wind. Such filaments have been both detected in real winds, such as M82 \citep[][]{Shopbell98} and NGC839 \citep[][]{Rich10}, or predicted in theoretical simulations \citep[][]{Cooper08}.  Their detection in the wind of NGC838 will require further observations with higher spatial resolution to be confirmed. We recall that the Southern region was observed under slightly better seeing conditions (1.5"-2"), which may explain why no hints of a filamentary structure is found in the Northern region.

In this toy model, the nature of the South East Clump described in Sec.~\ref{sec:velstruct} remains somewhat of a mystery. \cite{Cooper08} mention that in their simulations, the H$\alpha$ filaments resulting from the breakup of gas clouds in the wind do not trace the true extent of the hot gas (defining the hot bubble). The South East Clump may therefore simply be the sign that the Southern wind bubble is more extended, and its breakout more generalised than shown in our toy model (see Figure~\ref{fig:schema}). Another possible explanation, recalling that the signature of A-type stars is detected throughout the disk of NGC838, is that this H$\alpha$-bright clump is a remnant of another, earlier outflow episode related to a more global starburst. This over-density clump would become visible as it is photoionized by the central starburst in NGC838. Given its spatial location and blueshifted radial velocity of -122$^{\pm6}$ km s$^{-1}$, the South East Clump does not seem to be related to the current Southern wind bubble and its highly redshifted material. A careful inspection of the SINGG H$\alpha$ image of NGC838 reveals other knots of emission further to the East, outside of our WiFeS observations. Follow-up observations are required to measure the radial velocity of these H$\alpha$ knots, and, if confirmed, identify their nature (e.g. wind clumps or tidal debris fragments) and the origin of their eventual blueshift. 

What triggered the initial starburst in NGC838 remains to be clarified. At this point, the most likely explanation is that tidal perturbations resulting from the interactions of NCG838 with its neighbour NGC839 triggered a gas infall fuelling the starburst. One could imagine another scenario \citep[supported by the H{\footnotesize I} map of HCG16, see][]{Verdes-Montenegro01} in which NGC848, colliding with HCG16 and passing close from NGC838 and NGC839, would trigger their respective nuclear starburst and galactic wind formation. NGC848 is located $\sim$14.2 arcminutes (204.5 kpc) to the South East of NGC838 (on-sky distance). Assuming an on-sky velocity of 400 km s$^{-1}$ for the intruder galaxy (two times the typical HCGs velocity dispersion), the closest passage would have occurred $\sim$500 Gyr ago. This is in good agreement with the age of the A-type star population detected throughout the disk. The Northern wind bubble is extending roughly 5 kpc (on-sky) out of the galaxy plane, which according to \cite{Chatt12} implies a wind age of 37 Myr. Following the simulations of \cite{Cooper08}, the Southern wind bubble gas velocities lower than 400 km s$^{-1}$ would indicate an even younger age for the wind, of the order of 1-2 Myr. The ratio of the wind bubble radius to its expansion velocity provides an alternative estimate of the wind expansion timescale. For the Northern bubble with a radius of $\sim$5 kpc (on-sky) and an expansion velocity of $\sim$100 km s$^{-1}$, we obtain an age of 49 Myr. For the Southern bubble with a radius of 7 kpc (on-sky) and an expansion velocity of 400 km s$^{-1}$, we obtain an age of 17 Myr.   
These ages are only rough estimates \citep[see, e.g.][and references therein for a more detailed treatment of the dynamics of galactic winds]{Koo92a,Koo92b,Veilleux05}. There is a known delay \citep[$\geq$10Myr, see][]{Sharp10} between a starburst episode and the onset of a galactic wind. Yet, the 2-3 orders of magnitude discrepancy between our first-order wind age estimates and the closest approach between NGC838 and NGC848 makes the latter a somewhat unlikely \emph{direct} trigger for the activity seen in NGC838 today. It is plausible that a close encounter with NGC848 triggered a global starburst episode in NGC838, which has since then settled down to the most central regions of the galaxy. The South East Clump may be one of the signature of this past interaction between NGC838 and NGC848. 

Line ratio diagnostics indicate that the base of the wind is photoionized by the central starburst, and that mixing with slow shocks increases further out along the outflow. Mixing is mostly detected in the Northern wind bubble, which is consistent with it being still well contained with its surrounding H{\footnotesize I} envelope. \cite{Sharp10} noted in IFS observations of 10 low-redshift galactic winds that shock excitation usually dominates in instantaneous starburst driven wind. The largely photoionized starburst-driven wind of NGC838 therefore argues in favour of our long-lasting, multi-phase starburst scenario (initially global, now central).

\subsection{NGC838 as a starburst-driven galactic wind}

Emission line splitting of a few hundred km s$^{-1}$ is a typical signature associated with many galactic winds \citep[see, e.g.][]{Heckman90,Veilleux05,Bland07,Sharp10}. The classical interpretation is that the emitting gas is located on the surface of an open-ended bubble or conical structure. In NGC838,  while we detect multiple kinematic components in the emission line profiles, a clear split is not evident in the Northern region. This result indicates that the Northern wind bubble is still contained within its surroun envelope, and its interior not yet freely outflowing in the galaxy halo. NGC838 Northern bubble is therefore caught early on in the galactic wind evolution sequence, a few Myr only after the wind starts to blow \citep[][]{Veilleux05,Cooper08}.

Large morphological differences exist between different galactic winds \citep[see][for an extensive review]{Veilleux05,Bland07}. These differences are related to the wind excitation mechanism (starburst or AGN), the host galaxy structure, and the amount of gas present in the system. In the case of NGC838, the prime factor giving rise to the peculiar, asymmetric morphology of the outflow is that this wind is seen early on in its evolution.

Limb-brightening is another feature seen in several galactic winds, for example in NGC 3079 \citep[][]{Veilleux94}, Arp 220 \citep[][]{Colina04}, NGC 253 \citep[][]{Matsubayashi09} or the Milky Way \citep[][]{Bland03}. In NGC838, we do not see any clear evidence for limb-brightening, suggesting that the wind bubbles are not hollow, but rather filled with the emitting gas. It is understood that optical emission in galactic winds originate from disk material entrained in the outflow, an interpretation confirmed by theoretical simulations \citep[][]{Cooper08, Cooper09}. The clear rotation signature that we detect in the Northern wind bubble of NGC838 is consistent with this entrained origin of the emitting gas. We note here that the wind in NGC838 is most certainly very dissimilar to the limb-darkened wind in the Circinius galaxy \citep[][]{Veilleux97, Sharp10}, which is a Seyfert 2 galaxy with a much more evolved wind.

\subsection{Comparison with X-ray observations}
X-ray emission from HCG16 has been reported by many authors in the past \citep[e.g.][]{Saracco95,Ponman96,Turner01, Belsole03}. Most recently, \cite{Desjardins13} presented a new analysis of archival \emph{Chandra X-ray Observatory} \citep[][]{Weisskopf02} observations of HCG16 (ObsID: 923, P.I.: Mamon) as part of a larger study of diffuse X-ray emission in CGs. They provide emission maps of the diffuse soft X-ray emission (0.2-2.0 keV) in HCG16 which is clearly concentrated around the galaxy members of the group. The spatial correlation between the soft X-ray emission and the H$\alpha$ emission is evident for both NGC838 and NGC839 in the color mosaic of the group (combining R-band, H$\alpha$ and soft X-ray images). Keeping in mind that \cite{Desjardins13} have smoothed their Chandra data, we note that the soft X-ray emission associated with NGC838 is surrounding the H$\alpha$ emission by $\sim$5-10" in all directions except to the South-East, where the X-ray emission vanishes more rapidly. 

This lack of soft X-ray emission is consistent with the leaking Southern wind bubble scenario described previously. The low density and high temperature of the gas composing galactic winds is hard to detect directly \citep[see, e.g.][and references therein]{Bland07}. Instead, galactic winds are usually seen indirectly through photoionized, shock-excited or absorbing neutral material entrained from the disc. In NGC838, the wind in the Northern bubble has not yet burst out of the surrounding gaseous envelope. The wind's encounter with the surrounding material at the bubble's edge drives 1) a forward shock wave in the surrounding gaseous material, and 2) a reverse shock wave in the galactic wind. With velocities of the order of several hundred km s$^{-1}$ (i.e. of the order of the outflow velocity at the base of the wind), the reverse shock wave excites the wind material to X-ray temperatures which then cools via soft X-ray emission. The shocks will be brightest when the galactic wind is pushing through a denser surrounding environment, as the plasma emissivity is a function of the plasma density \citep[the energy lost via two-body interactions scales approximatively as the density squared. See, e.g.,][p.143-144]{Dopita03}. If the wind material can leak into a low-density region of the surrounding medium, as is likely the case for the Southern wind bubble of NGC838, then the shock-heated X-ray plasma would rapidly expand. This expansion will lead to a decrease in the density and emissivity of the plasma, that would then become too faint to be detected. 

\subsection{NGC838 versus NGC839}

The circumstances required to trigger galactic winds are not yet fully understood. Identifying the necessary ingredients and actions resulting in a large scale galactic wind is complicated by the fact that both the galaxies content and their environment are likely to play a role in the process. From that point of view, the galactic winds in NGC838 and NGC839 may represent a unique opportunity to improve our understanding of the formation mechanisms of galactic winds as a function of the galaxy characteristics, such as the disk thickness or the location of the starburst \citep[][]{Cooper08}.   

Located in the same region of the same CG, NGC838 and NGC839 are subject to very similar environments and interaction histories (including with the intruder galaxy NGC848). The on-sky disposition of galaxies within HCG16 and its H{\footnotesize I} distribution \citep[see the VLA H{\footnotesize I} map of][]{Verdes-Montenegro01} suggest that NGC838 and NGC839 are likely to be their respective closest interacting partner. Any difference in the galactic wind located in these galaxies must therefore be linked to intrinsic differences in the galaxies themselves. The asymmetric, photoionized wind of NGC838 is in strong contrast with the symmetric, shock-excited wind in NGC839 \citep[][see also Figure~\ref{fig:singg}]{Rich10}. Furthermore, the conical morphology with constant outflow velocity inferred by \cite{Rich10} for the wind in NGC839 is very different from the wind morphology in NGC838. Understanding the differences between NGC838 and NGC839 is therefore key to understanding the differences in their galactic winds. 

For both galaxies, resolving the star formation history and identifying the underlying stellar population needs to be addressed carefully. For example, it may be that NGC838 and NGC839 have experienced episodic, out-of-phase starburst events giving rise to the differences in their galactic winds. Confirming or disproving this scenario requires high S/N measurements of the stellar continuum. High resolution H{\footnotesize I} maps would also be beneficial to understand where the gas reserve is located and what is the exact extent of the interaction between NGC838 and NGC839.  

\section{Conclusion}\label{sec:conclusion}

We presented our IFS observation of NGC838 and its galactic wind. These are the first results of a series of observations targeting star forming galaxies in Compact Groups with the WiFeS instrument on the ANU 2.3m telescope at Siding Spring Observatory. Our observations reveal the complex signature of an asymmetric galactic wind in NGC838. Emission line ratio maps and diagnostic diagrams show that photoionization is the main excitation mechanism at the base of the wind, with mixing from slow shocks (up to 20\%) increasing away from the galaxy center along the outflow axis. The free flowing gas in an open-ended Southern bubble compared to the contained and rotating gas in a closed Northern bubble gives rise to the asymmetry of the wind. The closed Northern bubble is a strong indicator that the Northern wind is caught early (a few Myr) in the galactic wind evolution sequence. We also find kinematic evidence supporting a scenario where the Northern outflow is blocked and/or redirected at its base by thick dust lanes present in NGC838.

The presence of A-type stars throughout NGC838 suggests that the galaxy has been subject to a global episode of star formation some 500 Myr ago. The photoionized nature of the wind is consistent with this non-instantaneous starburst scenario, which started globally and has now settled down in the center of the galaxy. The H{\footnotesize I} map of HCG16 suggests that the intruder galaxy NGC848 may have triggered the original galaxy-wide starburst, as it collided with HCG16 and flew by NGC838.

Finally, the photoionized, expanding, rotating, asymmetric wind of NGC838 is in strong contrast with the symmetric and shock-excited wind of the neighbouring galaxy NGC839. Because they are subject to the same environment, and (most likely) the same interaction history, the differences in these two galactic winds are mostly due to intrinsic differences in their host galaxies. NGC838 and NGC839 therefore represent a unique pair of galaxies holding critical clues regarding the formation mechanisms of galactic winds.

\acknowledgments
We thank the anonymous referee for his/her constructive comments. M.A.D. acknowledges the support from the Australian Department of Science and Education (DEST) Systemic Infrastructure Initiative grant and from an Australian Research Council (ARC) Large Equipment Infrastructure Fund (LIEF) grant LE0775546 which together made possible the construction of the WiFeS instrument. M.A.D also acknowledges ARC support under Discovery project DP0984657. L.J.K. acknowledges the support of an ARC Future Fellowship and ARC discovery project grant DP130104879. This research has made use of NASA's Astrophysics Data System, NASA/IPAC Extragalactic Database (NED) which is operated by the Jet Propulsion Laboratory, California Institute of Technology, under contract with the National Aeronautics and Space Administration, and the Hubble Legacy Archive, which is a collaboration between the Space Telescope Science Institute (STScI/NASA), the Space Telescope European Coordinating Facility (ST-ECF/ESA) and the Canadian Astronomy Data Centre (CADC/NRC/CSA). Funding for SDSS-III has been provided by the Alfred P. Sloan Foundation, the Participating Institutions, the National Science Foundation, and the U.S. Department of Energy Office of Science. The SDSS-III web site is http://www.sdss3.org/. SDSS-III is managed by the Astrophysical Research Consortium for the Participating Institutions of the SDSS-III Collaboration including the University of Arizona, the Brazilian Participation Group, Brookhaven National Laboratory, University of Cambridge, Carnegie Mellon University, University of Florida, the French Participation Group, the German Participation Group, Harvard University, the Instituto de Astrofisica de Canarias, the Michigan State/Notre Dame/JINA Participation Group, Johns Hopkins University, Lawrence Berkeley National Laboratory, Max Planck Institute for Astrophysics, Max Planck Institute for Extraterrestrial Physics, New Mexico State University, New York University, Ohio State University, Pennsylvania State University, University of Portsmouth, Princeton University, the Spanish Participation Group, University of Tokyo, University of Utah, Vanderbilt University, University of Virginia, University of Washington, and Yale University. 

\appendix

\section{Dust reddening corrections}
There exist many ways to correct for extragalactic reddening, which is in itself not very well defined. Unfortunately, the literature is rather confusing, and it is often left to the reader to reconstruct by himself the exact solutions implemented in a given paper. Many authors also confuse \emph{extinction}, which applies along a single sight line, with \emph{mean attenuation}, which is a statistical sum of the attenuations produced by a foreground dust screen with a fractal distribution \citep[see the papers by][for the mathematical treatment of this]{Fischera05,Fischera11}. 

Here, we base our extragalactic reddening corrections on the work of \cite{Fischera05}, which \cite{Wijesinghe11} have shown provides the best agreement between star formation rate indicators in GAMA galaxies \citep[][]{Driver09}. The model assumes that the attenuation is caused by a distant isothermal and turbulent dust screen. \cite{Fischera05} provide several theoretical estimates of the relative attenuation as a function of wavelength which allow to correct a given spectrum based on its associated H$\alpha$/H$\beta$ ratio. \\

For clarity, let us derive the correction equation applied to our observations, starting from first principles. Let us adopt the formalism of \cite{Osterbrock89} (see their Chater~7). Given a luminous source subject to attenuation, we define $F_{\lambda}$ the observed flux at a given wavelength $\lambda$ (in the source reference frame), and $F_{\lambda,0}$ the theoretical, un-reddened flux. Then, 
\begin{equation}\label{eq:flux}
F_\lambda = F_{\lambda,0}\cdot e^{-\tau_\lambda},
\end{equation}
where $\tau_\lambda$ is the opacity along the object line-of-sight. Both scattering and absorption of the object light by the intervening dust are likely to play a role, and in fact, $\tau_\lambda=\tau_{\lambda,abs}+\tau_{\lambda,sca}$ \citep[see, e.g.][p.296]{Dopita03}. With $m_\lambda$ and $m_{\lambda,0}$ the observed and theoretical, un-reddened magnitude of the object, the extinction on the line-of-sight $A_\lambda$ for a given wavelength (in magnitude) is 
\begin{equation}\label{eq:alambda}
A_\lambda = m_\lambda-m_{\lambda,0}=2.5\log(e)\tau_\lambda.
\end{equation}
The actual shape of the opacity curve is highly debated. It is generally accepted as a simplification and based on galactic observations that $\tau_\lambda$ can be written as
\begin{equation}\label{eq:tau}
\tau_\lambda = c\cdot f_\lambda,
\end{equation}
where the \emph{shape} of the curve is described by a universal function $f_\lambda$, and $c$ is a multiplicative constant varying for every line-of-sight. Determining $c$ can be done by comparing the observed ratio of two given lines to the theoretically expected value, for example the H$\alpha$ and H$\beta$ lines. Using Eq.~\ref{eq:flux} and Eq.~\ref{eq:tau}, we have
\begin{equation}
\frac{F_{H\alpha}}{F_{H\beta}}=\frac{F_{H\alpha,0}}{F_{H\beta,0}}\cdot e^{-c(f_{H\alpha}-f_{H\beta})},
\end{equation}
and re-arranging the terms, 
\begin{equation}
c=-(f_{H\alpha}-f_{H\beta})^{-1}\cdot(\log(e))^{-1}\cdot \log\left(\frac{F_{H\alpha}/F_{H\beta}}{F_{H\alpha,0}/F_{H\beta,0}}\right).
\end{equation}
Assuming Case B recombination, $F_{H\alpha,0}/F_{H\beta,0}=2.85$. However, it has also been shown that regions close from active galactic nuclei may have a higher intrinsic ratio value of 3.1 \citep[e.g.][]{Kewley06}. \\

To find de-reddened flux value, we substitute the above expression for $c$ in Eq.~\ref{eq:flux}, which with simple algebra, and after re-arranging the terms slightly, becomes 
\begin{equation}\label{eq:flux_corr}
F_{\lambda,0} = F_\lambda\cdot \left(\frac{F_{H\alpha}/F_{H\beta}}{F_{H\alpha,0}/F_{H\beta,0}}\right)^{-f_\lambda/(f_{H\alpha}-f_{H\beta})},
\end{equation}
where $f_\lambda$ is the only unknown. Following \cite{Fischera05}, we can write 
\begin{equation}\label{eq:E}
\frac{E_{\lambda-V}}{E_{B-V}}=\frac{A_\lambda-A_V}{A_B-A_V},
\end{equation}
where $E_{\lambda-V}$ is usually refereed to as the \emph{color excess} between two bands, and $E_{\lambda-V}/E_{B-V}$ is the \emph{relative color excess}. Introducing $R_V^A=A_V/E_{B-V}$, Eq.~\ref{eq:E} becomes
\begin{equation}
A_\lambda = \left(\frac{E_{\lambda-V}}{E_{B-V}}+R_V^A\right)\cdot E_{B-V}.
\end{equation}
Therefore, 
\begin{equation}\label{eq:f}
\frac{f_\lambda}{f_{H\alpha}-f_{H\beta}} = \frac{\tau_\lambda}{\tau_{H\alpha}-\tau_{H\beta}}=\frac{A_\lambda}{A_{H\alpha}-A_{H\beta}}=\frac{\frac{E_{\lambda-V}}{E_{B-V}}+R_V^A}{\frac{E_{H\alpha-V}}{E_{B-V}}-\frac{E_{H\beta-V}}{E_{B-V}}}.
\end{equation}
Finally, substituting Eq.~\ref{eq:f} in Eq.~\ref{eq:flux_corr}, we find the reddening correction function,
 \begin{equation}\label{eq:corr}
 F_{\lambda,0} = F_\lambda\cdot \left(\frac{F_{H\alpha}/F_{H\beta}}{F_{H\alpha,0}/F_{H\beta,0}}\right)^{ -\frac{\frac{E_{\lambda-V}}{E_{B-V}}+R_V^A}{\frac{E_{H\alpha-V}}{E_{B-V}}-\frac{E_{H\beta-V}}{E_{B-V}}}}.
 \end{equation}
 
\begin{figure}[htb!]
\centerline{\includegraphics[scale=0.35]{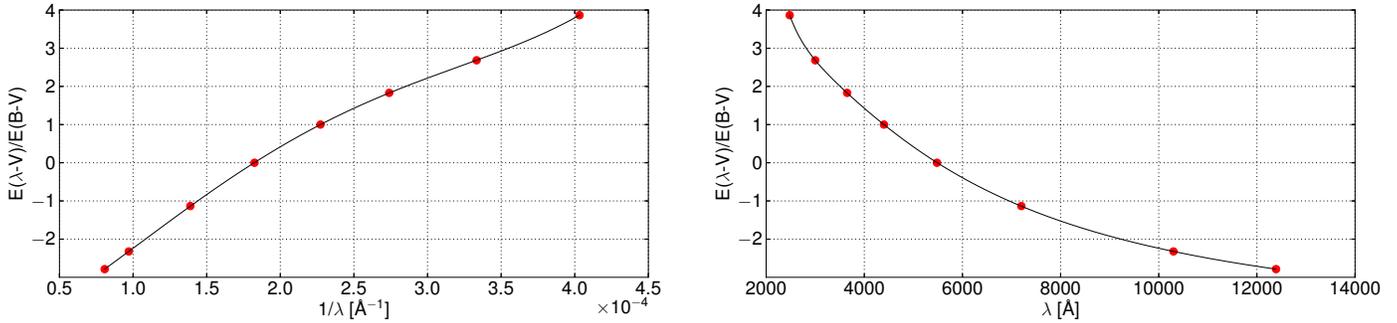}}
\caption{Relative attenuation curve from \cite{Fischera05} for $R_V^A=4.5$ and $A_V=1$ as a function of $1/\lambda$ (left) and $\lambda$ (right). The dots are from the Table~1 in the paper, and the black curve is our best fit 4$^{th}$ order polynomial, defined in Eq.~\ref{eq:poly}.}\label{fig:fischera}
\end{figure}

A value of $R_V^A$ = 4.3 results in an attenuation curve very similar to that defined empirically by \cite{Calzetti01} for starburst galaxies \citep[][]{Fischera03,Fischera05}. We therefore adopt the relative extinction curve with $R_V^A=4.5$ and $A_V=1$ of \cite{Fischera05}. From their Table~1, it is possible to extract the value of  $E_{\lambda-V}/E_{B-V}$ for various values of $\lambda$. From the few data points they provide in their paper, we derived a 4$^{th}$ order polynomial (using a least-square fitting routine) allowing us to obtain an accurate estimate of $E_{\lambda-V}/E_{B-V}$ for any value of $\lambda\in$[2480\AA;12390\AA]. The derived function is shown in Figure~\ref{fig:fischera} (black line) alongside with the points from \cite{Fischera05}, and can be written as 
\begin{equation}\label{eq:poly}
\frac{E_{\lambda-V}}{E_{B-V}} \cong -4.61777 + 1.41612\cdot\lambda^{-1} + 1.52077\cdot\lambda^{-2} - 0.63269 \cdot\lambda^{-3} + 0.07386\cdot\lambda^{-4}
\end{equation} 
with $\lambda$ in $\mu$m. We can then write Eq.~\ref{eq:corr} as
 \begin{equation}\label{eq:corr2}
 F_{\lambda,0} \cong F_\lambda\cdot \left(\frac{F_{H\alpha}/F_{H\beta}}{F_{H\alpha,0}/F_{H\beta,0}}\right)^{ 0.76\cdot\left(\frac{E_{\lambda-V}}{E_{B-V}}+4.5\right)}.
 \end{equation}

As a final remark, we note that we can also write the expression allowing to convert an H$\alpha$/H$\beta$ flux ratio to an $A_V$ magnitude, by substituting $c$ in Eq.~\ref{eq:alambda}, which gives 
\begin{equation}
A_V = -2.5 \log  \left(\frac{F_{H\alpha}/F_{H\beta}}{F_{H\alpha,0}/F_{H\beta,0}}\right) \cdot \frac{f_V}{f_{H\alpha}-f_{H\beta}}
\end{equation}
and using Eq.~\ref{eq:f},
\begin{equation}\label{eq:final}
A_V = -2.5 \log  \left(\frac{F_{H\alpha}/F_{H\beta}}{F_{H\alpha,0}/F_{H\beta,0}}\right) \cdot \frac{R_V^A}{\frac{E_{H\alpha-V}}{E_{B-V}}-\frac{E_{H\beta-V}}{E_{B-V}}}\cong8.55 \log  \left(\frac{F_{H\alpha}/F_{H\beta}}{F_{H\alpha,0}/F_{H\beta,0}}\right)
\end{equation}

\bibliographystyle{apj}
\bibliography{Vogt_2012}

\end{document}